\documentclass[aps,twocolumn,prb,showpacs,floatfix,superscriptaddress]{revtex4}
\usepackage{graphics,amssymb,amsmath,graphicx,subfigure}

\begin{document}

\title{Coarsening Dynamics of an Antiferromagnetic {\em XY} model on the Kagome Lattice:
Breakdown of the Critical Dynamic Scaling}

\author{Sangwoong Park}
\affiliation{Department of Physics,
The University of Suwon, Kyonggi-do 445-743, Korea}

\author{Bongsoo Kim}
\affiliation{Department of Physics,
Changwon National University, Changwon, 641-773, Korea}

\author{Sung Jong Lee}
\affiliation{Department of Physics,
The University of Suwon, Kyonggi-do 445-743, Korea}


\begin{abstract}

We find a breakdown of the critical dynamic scaling in the coarsening dynamics 
of an antiferromagnetic {\em XY} model on the kagome lattice when the system is quenched  
from disordered states into the Kosterlitz-Thouless ({\em KT}) phases at low temperatures.
There exist multiple growing length scales: the length scales of the average separation 
 between fractional vortices are found to be {\em not} proportional to the length scales of 
the quasi-ordered domains. They are instead 
related through a nontrivial power-law relation. 
The length scale of the quasi-ordered domains (as determined from optimal collapse 
of the correlation functions for the order parameter $\exp[3 i \theta (r)]$) 
does not follow a simple power law growth but exhibits an anomalous growth with 
time-dependent effective growth exponent. The breakdown of the critical dynamic 
scaling is accompanied by unusual relaxation dynamics in the decay of fractional 
($3\theta$) vortices, where the decay of the vortex numbers is characterized 
by an exponential function of logarithmic powers in time.

\end{abstract}



\maketitle

\today 

\section{Introduction}

Thermodynamic systems quenched from a high-temperature 
disordered phase into a low-temperature ordered phase exhibit characteristic growth 
of the length scale $\ell$ of the ordered domains which in typical situations 
can be represented as 
a power law  $\ell \sim t^{1/z}$, where the growth exponent $1/z$ depends on the 
dimension of the space and of the relevant order parameter as well as on the 
conserved or nonconserved nature of the order parameter.\cite{review}

Phase ordering dynamics (or coarsening process) is usually accompanied
by the annihilation and decay of the characteristic topological defects 
such as point vortices or domain walls which are generated in the initial 
disordered states. 
One of the most important notions in understanding and analyzing these coarsening 
processes is the so-called {\em dynamic scaling hypothesis}\cite{review} 
for the equal-time spatial correlation function of the order parameter, which 
is closely related to the observed 
self-similarity of the coarsening systems at different time instants. 

When the low temperature phase of the system is characterized by a quasi-long-range 
order with power law decay of the spatial correlation function of the
order parameter at equilibrium, then the dynamic scaling is generalized to 
the {\em critical dynamic scaling}. 
One of the well known examples is the ferromagnetic {\em XY} model\cite{berezin72,kt73} 
on the square lattice.        

However, it should be noted that this (critical) dynamic scaling hypothesis for the 
phase ordering (or quasi-ordering) dynamics has not been proved on some general theoretical 
basis. Therefore, it is not clearly known what is the condition for the validity of the 
dynamic scaling which is usually assumed in analyses of experimental results or numerical 
simulations on coarsening dynamics.

In typical situations, coarsening dynamics is investigated on
systems where the phase ordering is accompanied by the breaking of discrete
or continuous global symmetry. However, there exist model systems that exhibit
infinite ground state degeneracies of a discrete nature in addition to the usual 
(global) continuous symmetry. 
Prominent examples are geometrically frustrated spin systems such as antiferromagnetic 
Ising models on a triangular lattice\cite{uhkim2003,uhkim2007}, antiferromagnetic 
{\em XY} or Heisenberg models on kagome\cite{syozi51} or 
pyrochlore lattices.\cite{moessner2006} These systems exhibit interesting 
equilibrium and nonequilibrium behavior due to the geometric frustration effect, including 
spin-glass-like relaxation dynamics without quenched disorder.   

In this work we perform numerical simulations on the coarsening of 
an antiferromagnetic {\em XY} model on the kagome 
lattice ({\em KAFXY} model)\cite{huse92, rzchow97, cherepa2001, korshu2002,kolah2002} 
which is one of the simplest geometrically frustrated models with infinite ground state 
degeneracies.
Experimentally, this model can be realized in superconducting Josephson-junction 
arrays or superconducting wire networks\cite{higgins2000,park2001,xiao2002} on 
a kagome lattice 
when a perpendicular magnetic field of half flux quantum (per plaquette) is applied 
on the system.  We can also find physical examples in the anisotropic limit of 
Heisenberg antiferromagnets on the kagome lattice. It is well known that the system 
exhibits an infinite ground state degeneracy with finite entropy\cite{baxter70}. 
The system also exhibits a finite temperature 
{\em KT} transition\cite{kt73} corresponding to the unbinding of so-called 
fractional $3\theta$-vortices. That is, at low temperature below the {\em KT} transition, 
equilibrium of the system will be characterized by quasi-long-range order of the
order parameter $ \psi_3 \equiv e^{3 i \theta (r)}$. Analogous to the case of 
simple ferromagnetic {\em XY} model on a square lattice, we might expect that the coarsening
dynamics of the system would exhibit a critical dynamic scaling for the equal-time spatial 
correlation of the order parameter.

Our simulations, however, show that the critical dynamic scaling is not obeyed very well 
at least for the time duration of our numerical simulations ranging several decades 
of time scale. That is, it was not possible to achieve a good scaling collapse for
the equal-time spatial correlation functions of the $3\theta$ order parameter.     
Another signature of the breakdown of the critical dynamic scaling is that the length 
scales corresponding to the average separation between fractional vortices are  
not proportional to the length scales of the growing quasi-ordered domains, which 
are instead related through a nontrivial power-law relation. This means that the two 
length scales exhibit different growth behavior in time.   
In terms of the decay of the fractional vortices, we also found that the fractional
 $3\theta$-vortices residing in the small triangular plaquettes exhibit faster decay, 
while, in contrast, those vortices sitting on the larger hexagonal plaquettes exhibit 
much slower decay that can be fitted by an exponential of logarithmic powers in 
time\cite{cherepa94}.  

In addition to these features of multiple length scales in the coarsening dynamics,
the time dependence of the length scale of the quasi-ordered domains does not exhibit
a simple power law growth but rather exhibits an anomalous growth with time-dependent 
effective growth exponent. This appears to be closely related to the  unusual 
relaxation dynamics in the non-power-law decay of the fractional ($3\theta$) vortices.
It is not clear yet whether the scaling may be restored in the asymptotic 
limit.

\section{ The model system and simulation methods}

\begin{figure*}
\centering
\includegraphics[width=7cm]{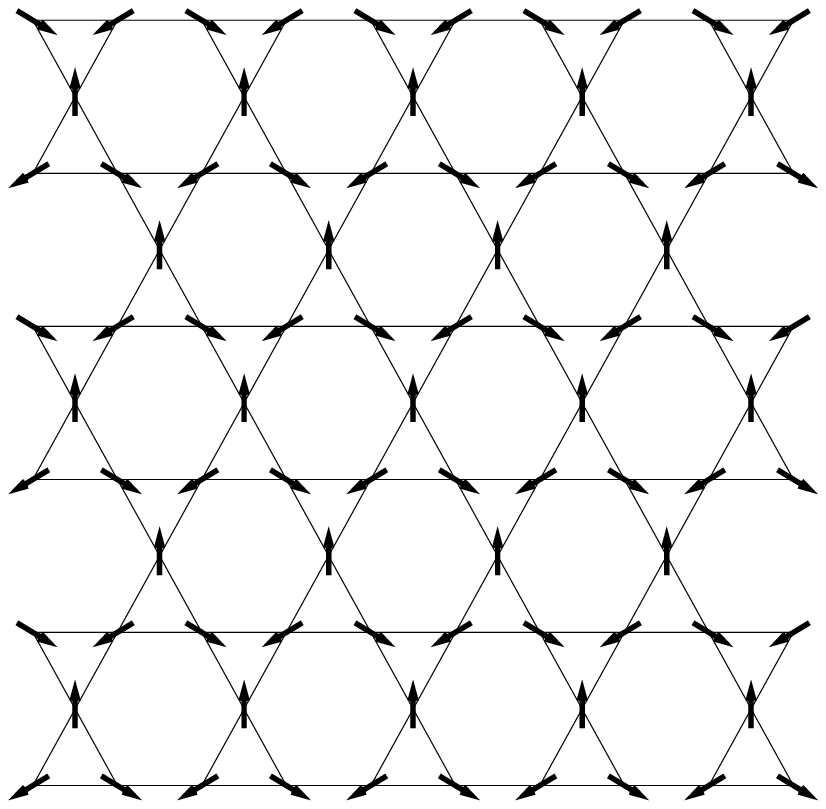}
\centerline{(a)}
\includegraphics[width=7cm]{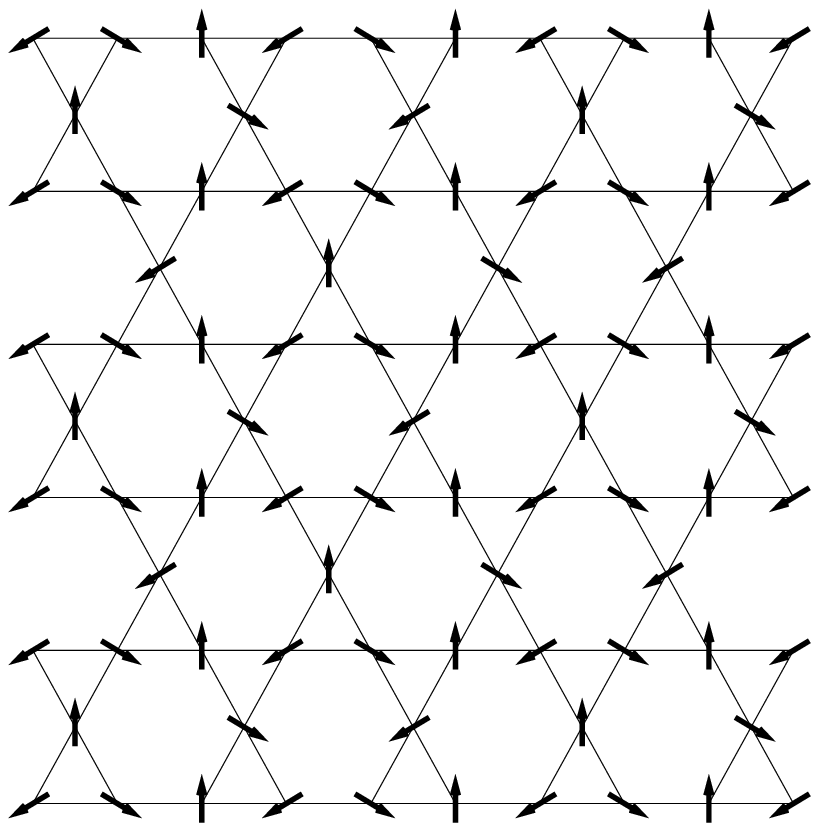}
\centerline{(b)}
\caption{ Schematic drawings of kagome lattices with the so-called 
(a) ${\bf q} = 0$ ground state and (b) ${\bf q} = \sqrt{3} \times \sqrt{3} $ ground state. } 
\label{kagome_L8}
\end{figure*}

The kagome lattice consists of corner-sharing triangles (Fig.~\ref{kagome_L8}). 
In an antiferromagnetic {\em XY} model on the kagome lattice, the 
Hamilitonian is defined as
\begin{equation} \label{KXY}
H = -J\sum_{\langle i,j\rangle} \cos(\theta_{i}-\theta_{j}).
\end{equation}
where $J<0$, the sum runs over all nearest neighbor pairs of sites, and $\theta_{i}$ 
denotes the angle of the planar spin at site $i$ with respect to some fixed direction in 
the two dimensional spin space. 
$\langle i,j\rangle$ indicates all pairs of nearest neighbor sites in the kagome lattice. 

It is easy to see that the ground states of this system have the property that 
for all pairs of nearest neighbors $i$ and $j$, the angle difference satisfies
$\left|\theta_{i}-\theta_{j}\right|= 2\pi/3$ which means that the sum of the three spins
vectors on any unit triangle vanishes.  
Thus the space of the ground states of {\em KAFXY} model are equivalent (up to a global rotation)
to the ground states of the three state Potts model on the kagome lattice.
In Figs.~\ref{kagome_L8}(a)-(b) are shown two examples of simple ground states with long-range 
order, the so-called ${\bf q} = 0$ state and ${\bf q} = \sqrt{3} \times \sqrt{3}$ state, where
${\bf q}$ refers to the wave vector corresponding to the periodicity of the chirality
configuration. It is easy to see, however, that in addition to these ground states 
with simple spatial order there also exist infinitely many ground states with no spatial order.      
It is well known that the system has a ground state entropy of $S_0 \simeq 0.126k_{B}$ 
per site\cite{baxter70}.  
Now, if we consider the angle variable $3\theta$ and the corresponding complex order parameter
$\Psi_3 \equiv \exp(i3\theta)$, the degenerate ground  states are all completely ordered in 
terms of this new  order parameter. And it has been shown that the system undergoes
a {\em KT} transition at a finite temperature, where the spatial correlation of the order 
parameter exhibits an algebraic decay below the transition temperature. 

As for the {\em KT} transition temperature of this system, an analytic approximation 
was given as \cite{cherepa2001,korshu2002} 
\begin{equation}\label{T_c}
T_c  = \frac{\pi \sqrt{3}}{72} J \approx 0.0756 J. 
\end{equation}
Numerical simulations were performed by Rzchowski\cite{rzchow97} where he found two 
slightly different estimates on the transition temperature, i.e., one estimate of 
$T_c  \simeq 0.076J$ 
based on the Binder cumulants of the order parameter and another of $T_c  \simeq 0.070J$ 
from the helicity modulus (or correspondingly the decay exponent $\eta(T_{c}) = 0.25$ 
of the spatial correlation of the order parameter). 

In the present work, the coarsening dynamics of the model system is performed via kinetic
 Monte Carlo methods with standard Metropolis algorithm for quenches to various 
temperatures near and below the {\em KT} transition. 
System sizes ranging up to $N \times N = 256\times 256$ were employed 
with periodic boundary conditions. In a kagome lattice, the number $\widetilde N $ 
of the total spins is  $\widetilde N = \frac{3}{4}N^2 $. 
The system is quenched from completely disordered initial states down to a given low
temperature with the process of coarsening being monitored through equal-time spatial 
correlation functions and the decay of topological defects, etc. 

In addition to quenches from disordered state to low temperatures, we also performed
the so-called nonequilibrium relaxation by suddenly bringing the system from one 
of the ground states to some target temperatures around or below the {\em KT} transition. 
This method was found to be convenient for measuring the values of the critical 
exponent $\eta$ for the equilibrium spatial correlation functions.

When we let the systems evolve from random initial configurations 
the following quantities can be measured:

\begin{enumerate}
\item The number density $n_v (t)$ of topological defects which are $3\theta$-vortices in 
      the {\em KAFXY} model    
\begin{equation}
n_v (t) \equiv \frac{\langle N_v(t) \rangle }{\tilde{N}} ,
\end{equation}
where $N_{v}(t)$ is the total number of $3\theta$-vortices (both positive and negative) 
at time $t$ and $\tilde{N}$ denotes the total number of sites (i.e., spins). 
$\left< \cdots \right>$ denotes an average over random initial configurations. 
We also count the separate number density of vortices residing on the 
hexagonal plaquettes ($n_h$) and those on the triangular plaquettes ($n_t$). 

\item The equal-time spatial correlation function of the $3\theta$ order parameter 
$\psi_{3}({\bf r}, t) \equiv e^{3i\theta({\bf r},t)}$.

\begin{eqnarray}
C(r,t) & = & \langle \psi_{3}^{\ast}({\bf r}, t)\, \psi_{3}(0,t)\rangle \\
      &   =  &  {1 \over {\widetilde N}} \left < \sum_{i}
        \exp (3i\theta_{i}(t) - 3 \theta_{i+r}(t)) \right>
\end{eqnarray}

\item Nonequilibrium spin autocorrelation functions
\begin{eqnarray}
A(t)  & = & {1 \over {\widetilde N}} \left < \sum_{i}
\exp (i\theta_{i}(0) - \theta_{i}(t)) \right>  \\
A_{3}(t)  &  =   &  {1 \over {\widetilde N}} \left < \sum_{i}
\exp (3i\theta_{i}(0) - 3\theta_{i}(t)) \right>.
\end{eqnarray}

\end{enumerate}

\section{simulation results}

\begin{figure*}
\centering
\includegraphics[width=8cm]{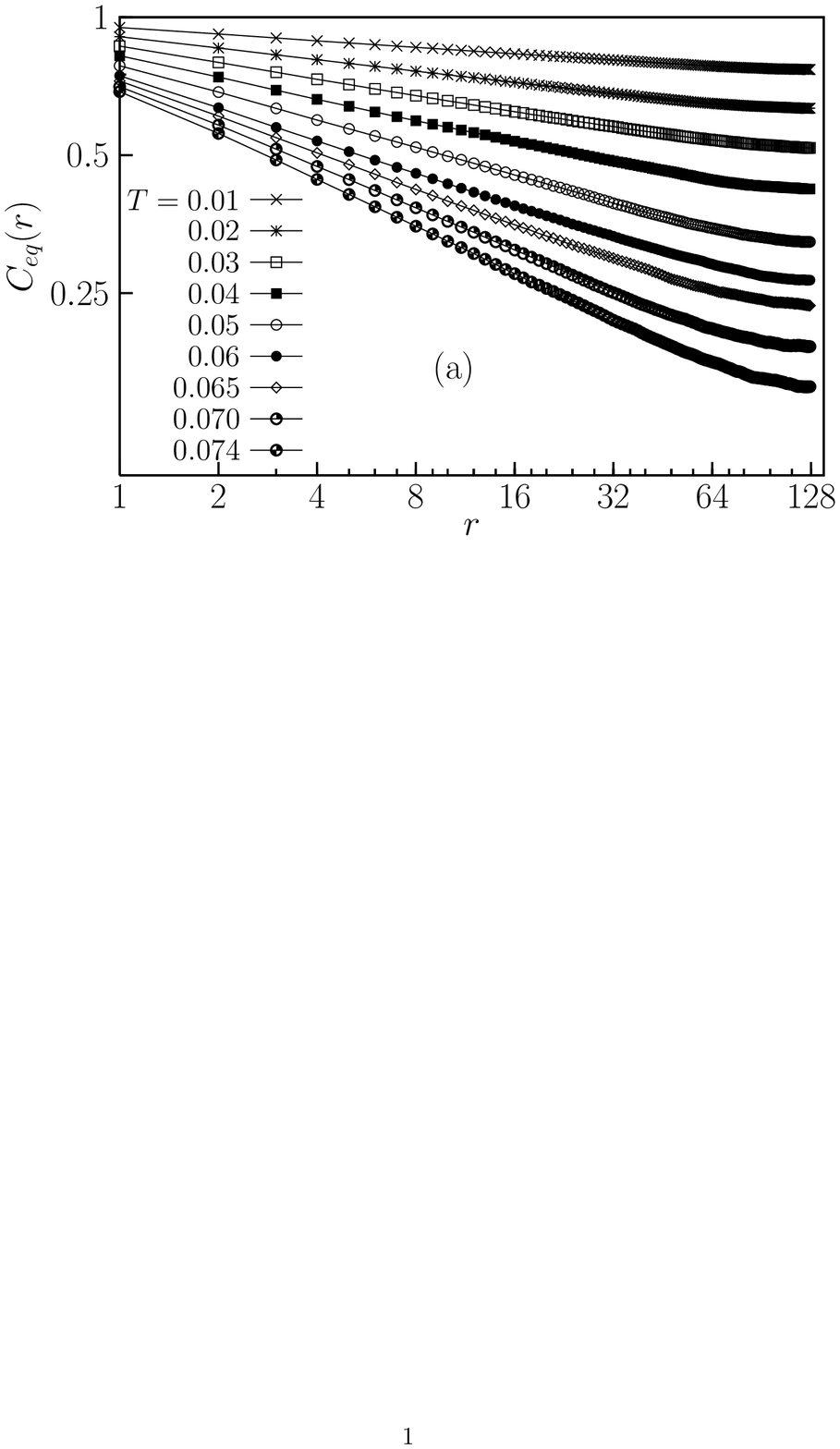}
\includegraphics[width=8cm]{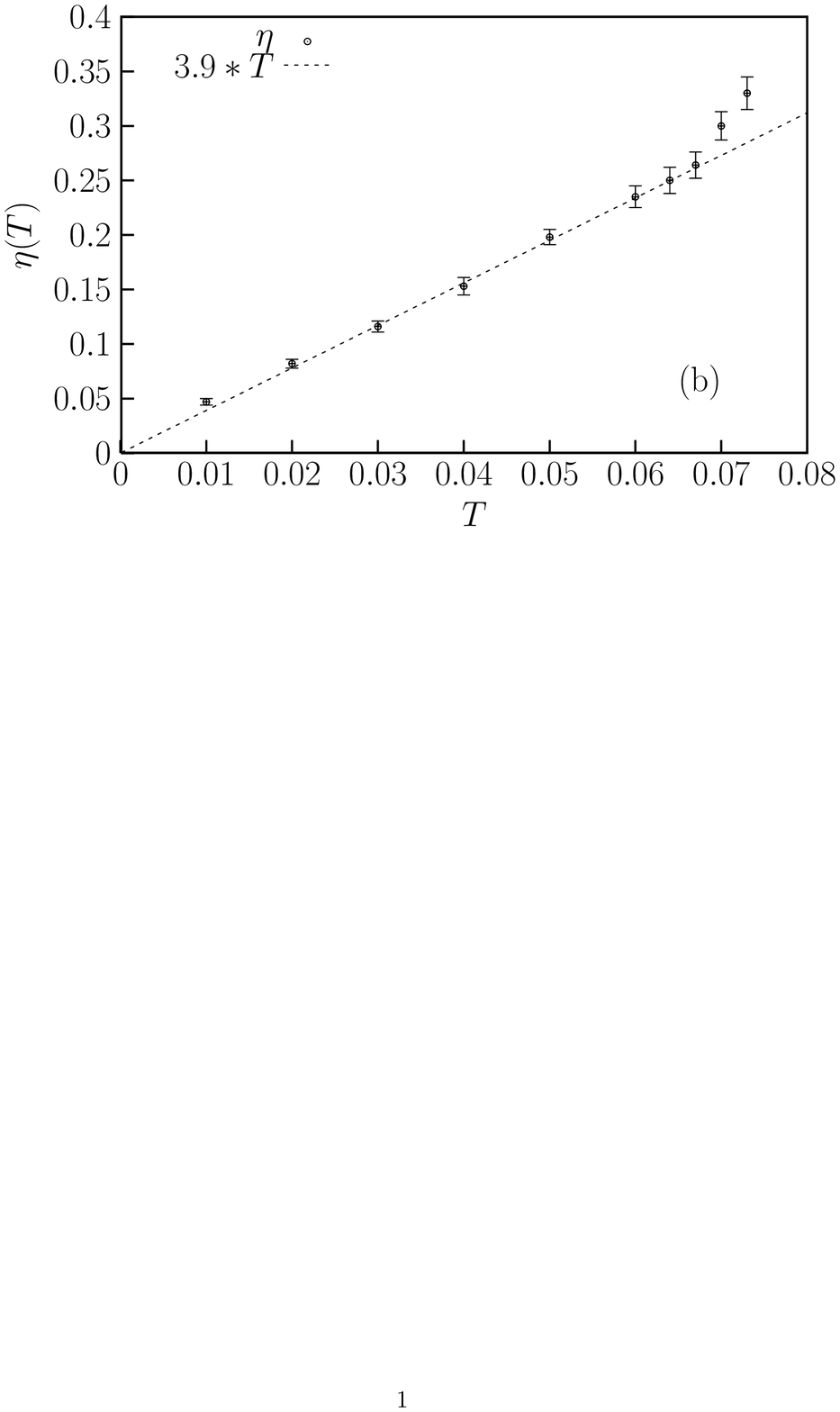}
\vspace*{0.5cm}
\caption{(a) Equilibrium spatial correlation of $\exp (3 i\theta)$ at various temperatures
     and (b) the corresponding exponents $\eta(T)$ vs. $T$, obtained from 
     nonequilibrium relxation method. In (b), the dotted line represent $3.9 T$ which 
     fits reasonably well the behavior of $\eta(T)$ vs. $T$ especially at low and intermediate 
     temperature regime ($T \leq 0.065$). } 
\label{spatial_corr_eta}
\end{figure*}

We have performed dynamic Monte Carlo simulations of {\em KAFXY} model on 
a kagome lattice of dimensions $256 \times 256$. In order to obtain the values of the 
equilibrium critical exponents $\eta $ for the spatial decay of the equilibrium spatial 
correlation function $C_{eq}(r)$ of the $3\theta$ order parameter, we have employed 
the so-called {\em nonequilibrium relaxation (NER)} method, where the system is 
suddenly brought from ground states to finite temperatures below or near the 
{\em KT} transition. 
Simulations were performed up to $655360$ Monte Carlo steps which was 
sufficient for the equilibrium to be attained. This was confirmed by the collapse 
of the spatial correlations at later time stages. 
In Fig.~\ref{spatial_corr_eta}(a) the correlation functions $C_{eq}(r)$ are displayed
on a log-log scale. The values of the exponent $\eta$ thus determined for temperatures 
ranging from $T=0.01$ to $T=0.074$ are shown in Fig.~\ref{spatial_corr_eta}(b) 
as well as in Table~\ref{t1}.

\begin{table}[h]\centering
\begin{tabular}{|c|c||c|c|} \hline \hline
      $T/J$ &  $\eta (T)$  &      $T/J$ &      $\eta (T)$  \\
\cline{1-4}
   0.01 &  0.047(3)  &  0.065  &    0.254(11)        \\
\cline{1-4}
  0.02 & 0.082(4) &     0.066  &    0.259(13)             \\
 \cline{1-4}
   0.03    & 0.116(5) & 0.067 &   0.264(12)    \\
 \cline{1-4}
   0.04  &  0.153(8)   & 0.068 &   0.268(12)       \\
\cline{1-4}
  0.05   & 0.198(7)    & 0.069  &  0.280(11)    \\
\cline{1-4}
 0.06 & 0.235(9) &    0.070  &     0.300(13)         \\
\cline{1-4}
 0.061 & 0.237(10) &   0.071  &    0.308(12)         \\
\cline{1-4}
 0.062 & 0.240(11) &  0.072  &    0.310(13)         \\
\cline{1-4}
 0.063 & 0.245(10) &  0.073  &    0.330(15)         \\
\cline{1-4}
 0.064 & 0.250(12) &  0.074  &    0.340(13)       \\
\hline\hline
\end{tabular}
\caption{The equilibrium exponent $\eta$ for different temperatures for the 
power-law decay of the spatial correlation of the $3\theta$ order parameter.}
\label{t1}
\end{table}

As for the generation of the initial low energy configurations,  we first select 
one of the three phases of $0$, $\pm 2\pi/3$ for each site at random and then
applied Monte Carlo procedure at very low temperature of $T=0.005$ for
$10000$ MC steps with the restriction that, in each of the MC procedure, a new trial 
phase is chosen only among the above three phases. In this way, the final states 
obtained were very close (in energy) to a ground state. These states were 
employed as the initial states of {\em NER} in order to bring the system to 
the equilibrium at some given target temperature.    

From the graph of $\eta (T)$ shown in Fig.~\ref{spatial_corr_eta}(b), we find that 
the exponent increases almost linearly in temperature up to around $T \simeq 0.065$, 
above which it starts to increase rather sharply. The linear regime can be fit 
approximately by $\eta (T) \simeq 3.9 T$. We also find that the exponent $\eta$ 
takes the value of $1/4$ near $T(\eta = 1/4 ) \simeq  0.064 J$ (Table~\ref{t1}).
 This temperature is expected to correspond to the {\em KT} transition, which apparently is 
a little lower than the theoretical\cite{cherepa2001,korshu2002} or the numerical\cite{rzchow97} 
estimates of earlier works mentioned above.

Now, the coarsening dynamics of the model system under quench to low temperature
from disordered initial states is investigated by monitoring the equal-time
spatial correlation functions of the $\psi_3$ order parameter.
One of the most important features of typical coarsening dynamics toward quasi-ordered
phase is the critical dynamic scaling
\begin{equation} \label{eq:dynamic_scaling}
C(r,t)= r^{-\eta(T)}f(r/L(t)),
\end{equation}
where $f(x)$ is the scaling function and $L(t)$ is the growing length scale.
It should be noted that the above scaling ansatz is based on the existence of a
single growing length scale $L(t)$.  

\begin{figure*}[htb]
\centering
\includegraphics[width=8cm]{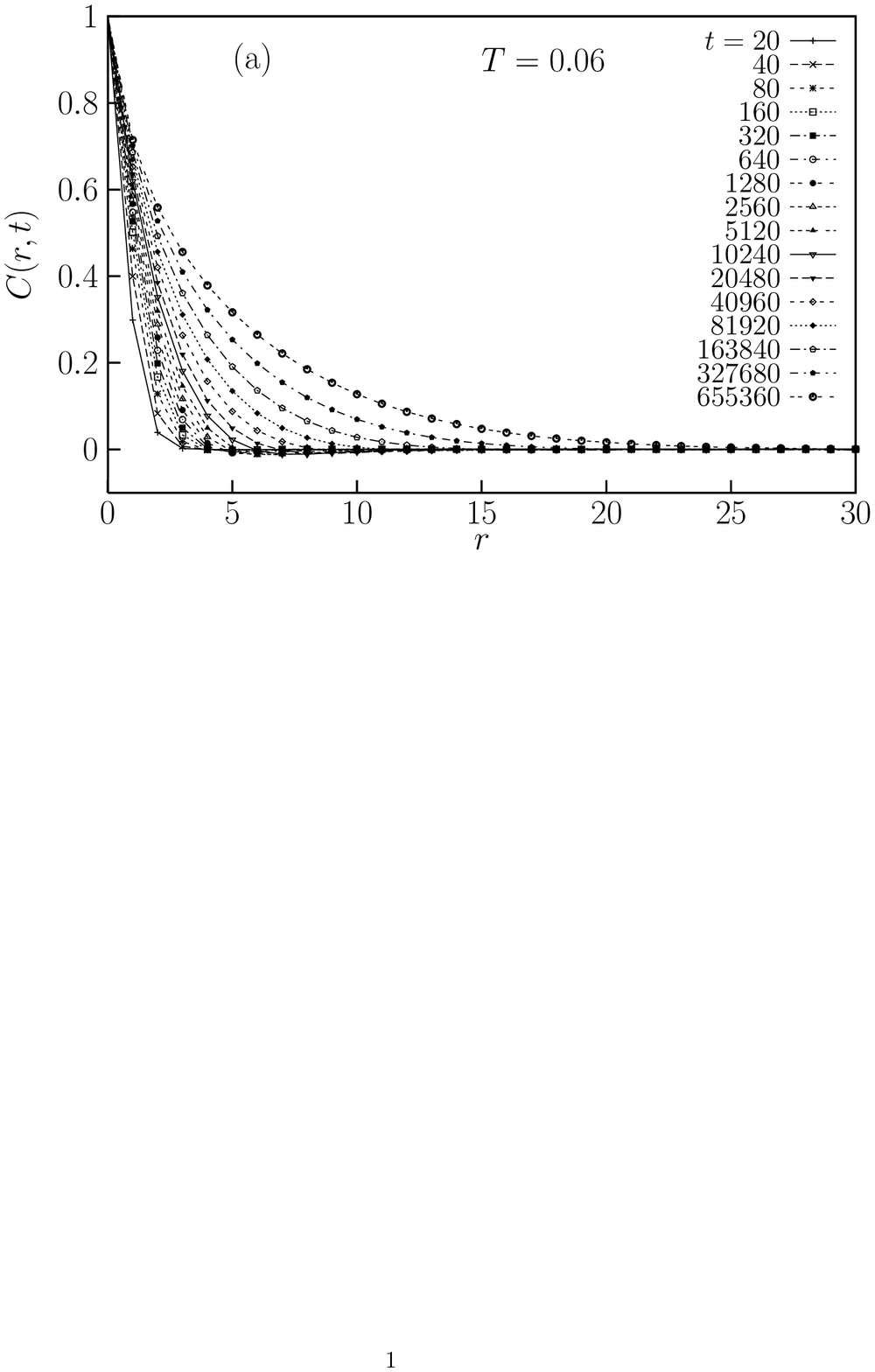}
\includegraphics[width=8cm]{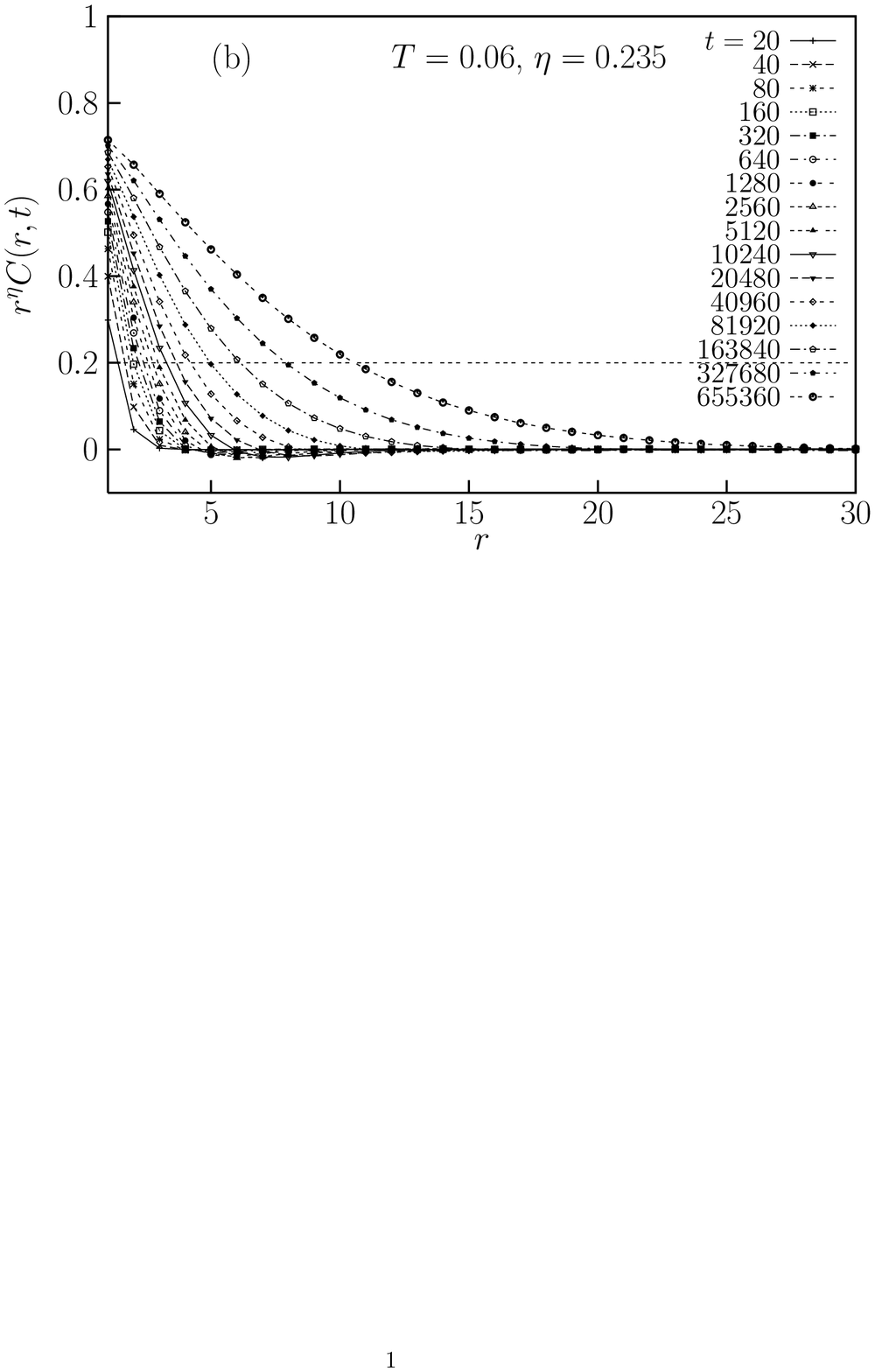}
\includegraphics[width=8cm]{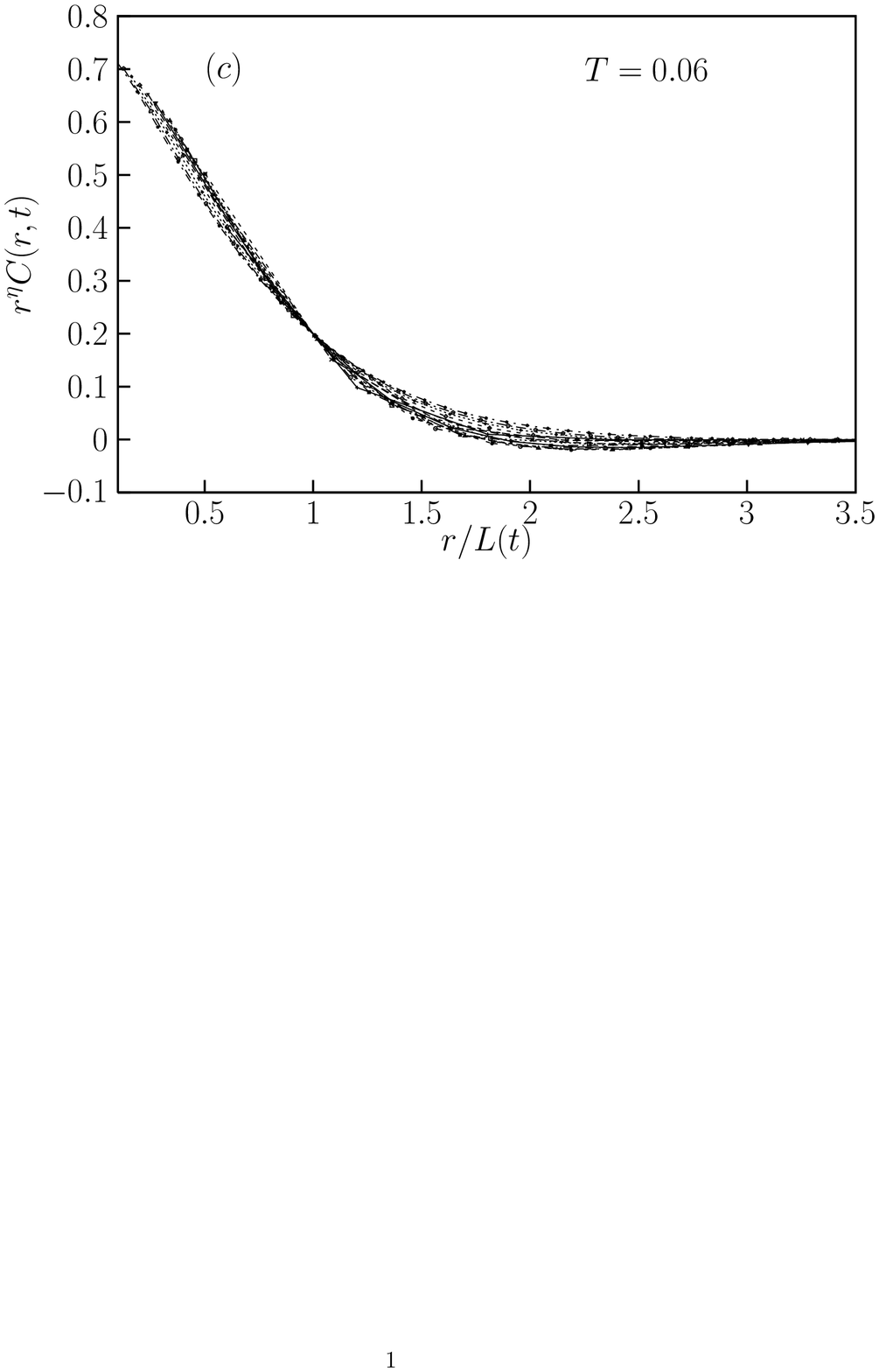}
\includegraphics[width=8cm]{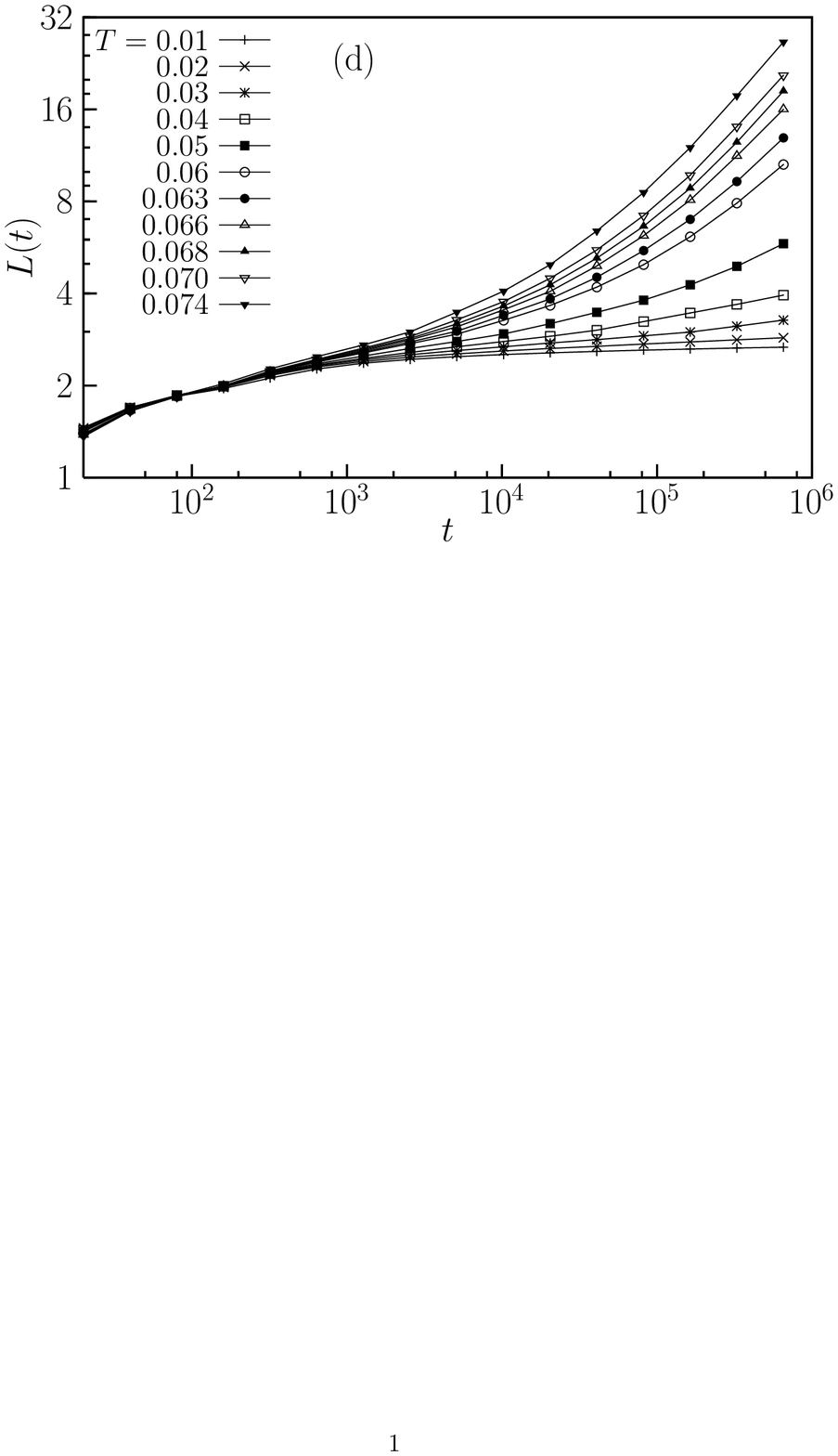}
\includegraphics[width=8cm]{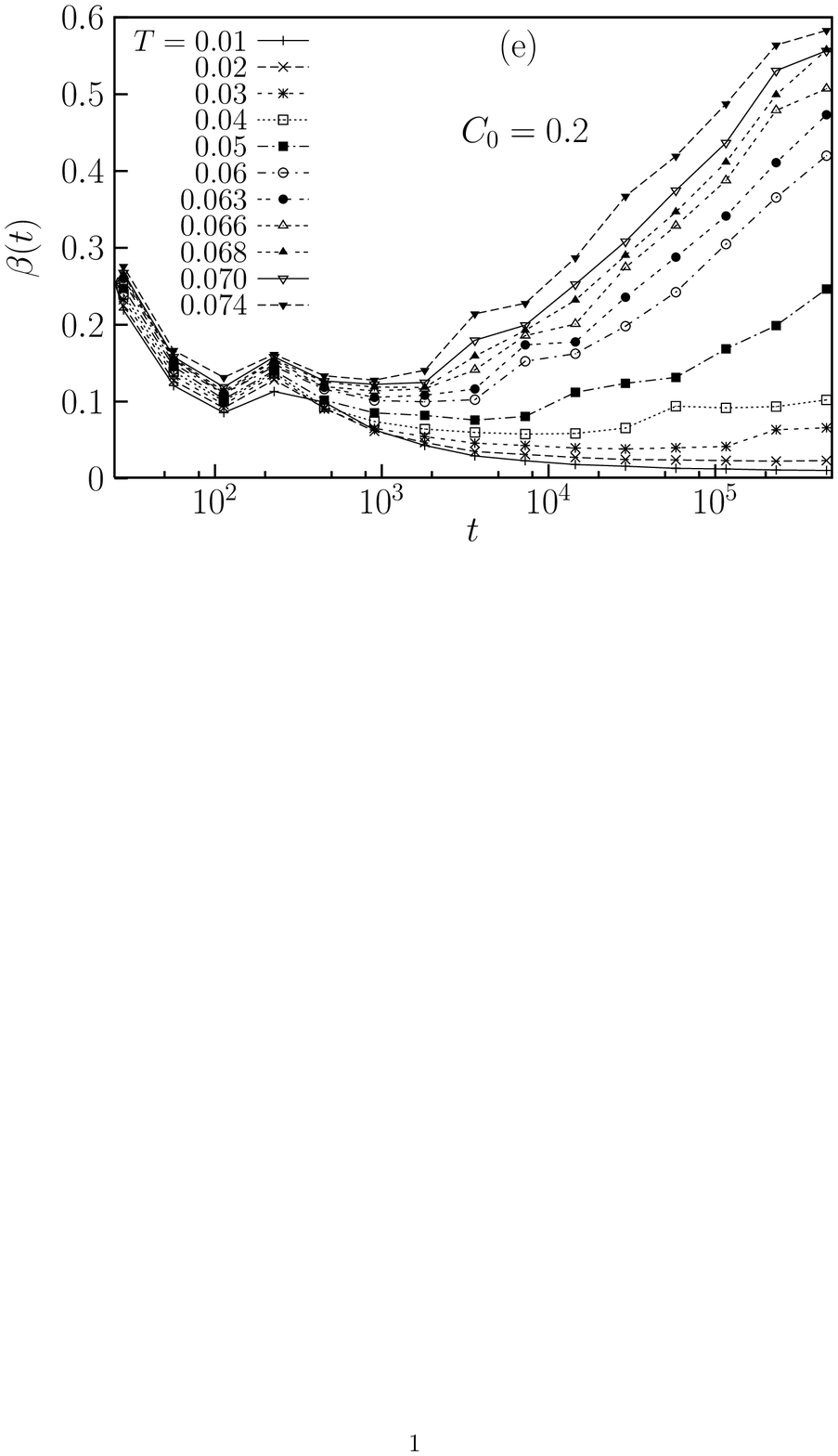}
\includegraphics[width=8cm]{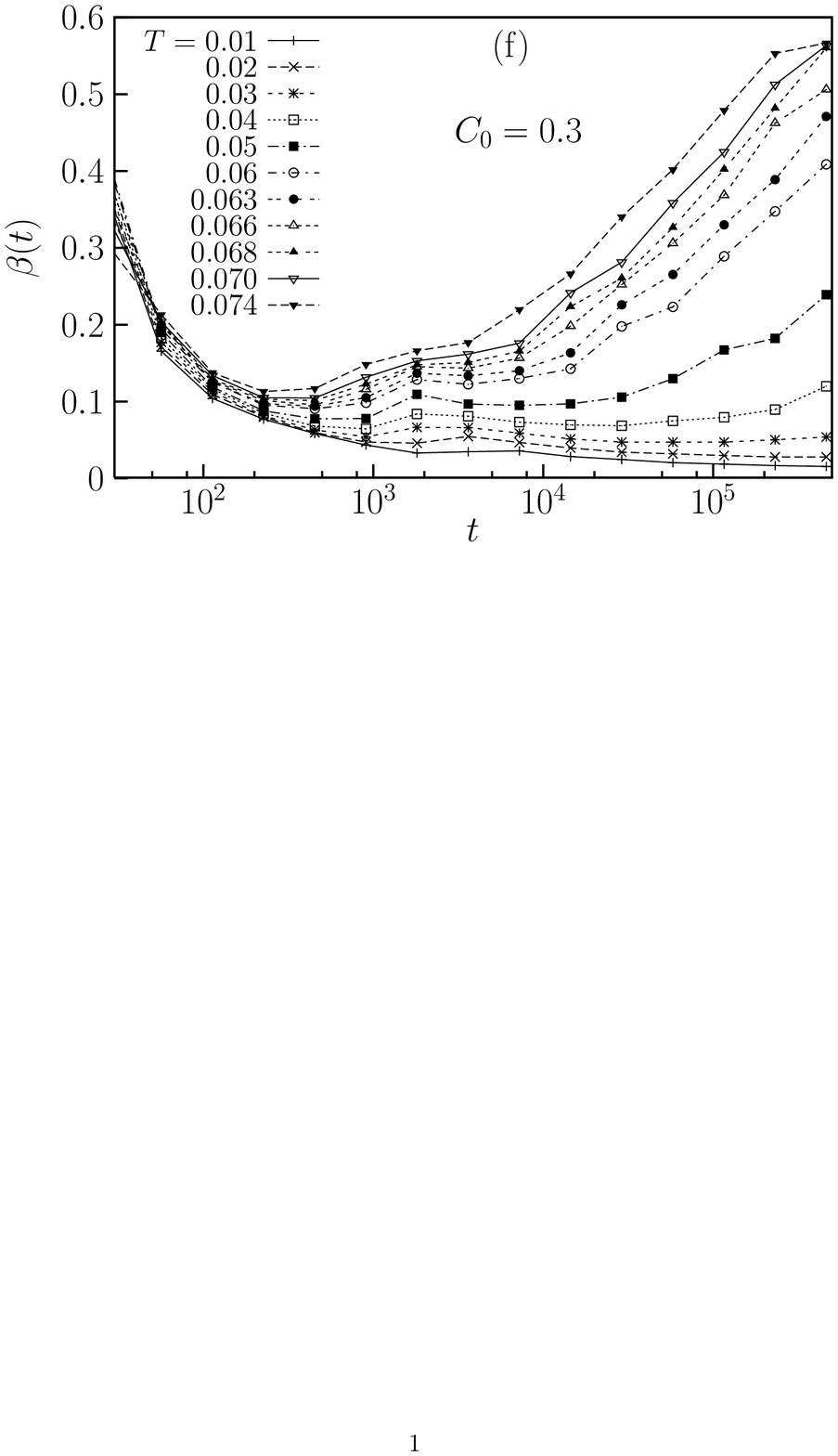}

\caption{(a) Equal-time spatial correlation function $C(r,t)$ for $\exp (3i\theta ) $ 
    for different time instants at $T=0.06$, (b) the rescaled functions 
    $\tilde{C}(r,t) \equiv r{^\eta(T) C(r,t) }$, (c) a scaling attempt based on (b)  
   with $C_0 = 0.2$, (d) growth of the length scale (using $C_0 = 0.2$), (e) the effective 
   growth exponents vs. time (using $C_0 =0.2$), and (f) the effective growth exponents vs. time
    (with $C_0 =0.3$).}
\label{spatial_corr_scaling}
\end{figure*}

Figure~\ref{spatial_corr_scaling}(a)-(b) shows the equal-time correlation functions 
$C(r,t)$ of {\em KAFXY} model for different time instants (from $t=10$ to $655360$) at 
the temperature $T=0.060$.  For these data, we attempted a critical dynamic scaling.
The procedure of rescaling is as follows. For a given temperature, we take  
the value of $\eta(T)$ (determined as in Table~\ref{t1}) and evaluate the combination 
$\widetilde C (r,t) \equiv r^{\eta (T)} C(r,t)$. 
And then, for a given time instant, the length scale $L(t)$ is determined by
the condition $\widetilde C(r=L(t),t) =  C_0 $ where the constant $C_0$ is chosen
as $C_0 =0.2$ or $C_0 =0.3$ in this work.  By plotting the scaled correlation
functions $\widetilde C(r,t)$ in terms of the rescaled distance $r/L(t)$, 
we can check whether the critical dynamic scaling holds or not from the quality 
of the collapse of the scaled correlation functions.  

By this procedure, we found, rather unexpectedly, that the critical dynamic scaling 
does not hold in the coarsening dynamics of {\em KAFXY} model, at least for the (Monte Carlo) 
time duration of our simulations (up to $655360$ MC steps). 
The result of the scaling attempt is shown in Fig.~\ref{spatial_corr_scaling}(c) for 
the case of $T=0.06$ where we can see that the critical dyanmic scaling is not obeyed. 
We also tried different values of $\eta$ for the rescaled correlation with no success. 
It might be possible that the scaling is restored in the limit of infinitely long time.

Even though the critical dynamic scaling is not faithfully obeyed, we may still 
extract approximate length scale $L(t)$ of spatial correlation in the manner described 
above. The length scale $L(t)$ thus obtained exhibits a rather unusual time-dependent 
behavior (Fig.~\ref{spatial_corr_scaling}(d)). That is, in the early-time stage up to 
around $t \sim 10^{3}$, $L(t)$ exhibits a slow growth, which appears approximately 
independent of the temperature. However, in the intermediate and late time stage of 
$t \gtrsim 10^{3}$, the growth of the length scale $L(t)$ becomes strongly  dependent 
on the temperature. In addition, for a given temperature, no simple power law is found
which is valid for the whole late time regime. Instead the local logarithmic slope
exhibits a steady increase in time in the late time regime.   

We can investigate the time-dependence of the growth of the length
scale by defining the effective local growth exponent as 
$\beta (t) \equiv d\ln(L(t)/d\ln(t)$.
Since we took the growing length scale $L(t)$ only for discrete time instants $t_i$ with
fixed interval in logarithmic scale, we evaluate the discrete version of the above
logaritmic slope as    
\begin{equation} \label{eq:effective_exponent}
\beta(t^{\prime}_{i}) \simeq \frac{ \ln (L(t_{i+1}) /L(t_i))}{\ln(t_{i+1}/t_i)}    
\end{equation}
where $t_i \equiv 2^{i}$, $i=1,2, \cdots$ and $t^{\prime}_{i} \equiv \sqrt{t_i t_{i+1}}$.
Figures.~\ref{spatial_corr_scaling}(e)-(f) show the effective growth exponent at time
$t$ (for the case of $C_0 = 0.2$ (Fig.~\ref{spatial_corr_scaling}(e)) and 
$C_0 = 0.3$ (Fig.~\ref{spatial_corr_scaling}(f)) respectively).
We can see that, in the late time regime, the growth becomes definitely faster as the 
temperature is higher. Also, for the temperatures $T \gtrsim 0.05$, the effective local
growth exponent appears to be monotonically increasing in the late time stage.   
At higher temperatures, we can recognize some indications from 
Figs.~\ref{spatial_corr_scaling}(e)-(f) that the effective slopes tend to converge 
around $0.5 \sim 0.6$ in the latest time regime. It may be possible to interpret this 
as an indication of an asymptotic behavior of diffusive growth.

\begin{figure*}[htb]
\centering
\includegraphics[width=8cm]{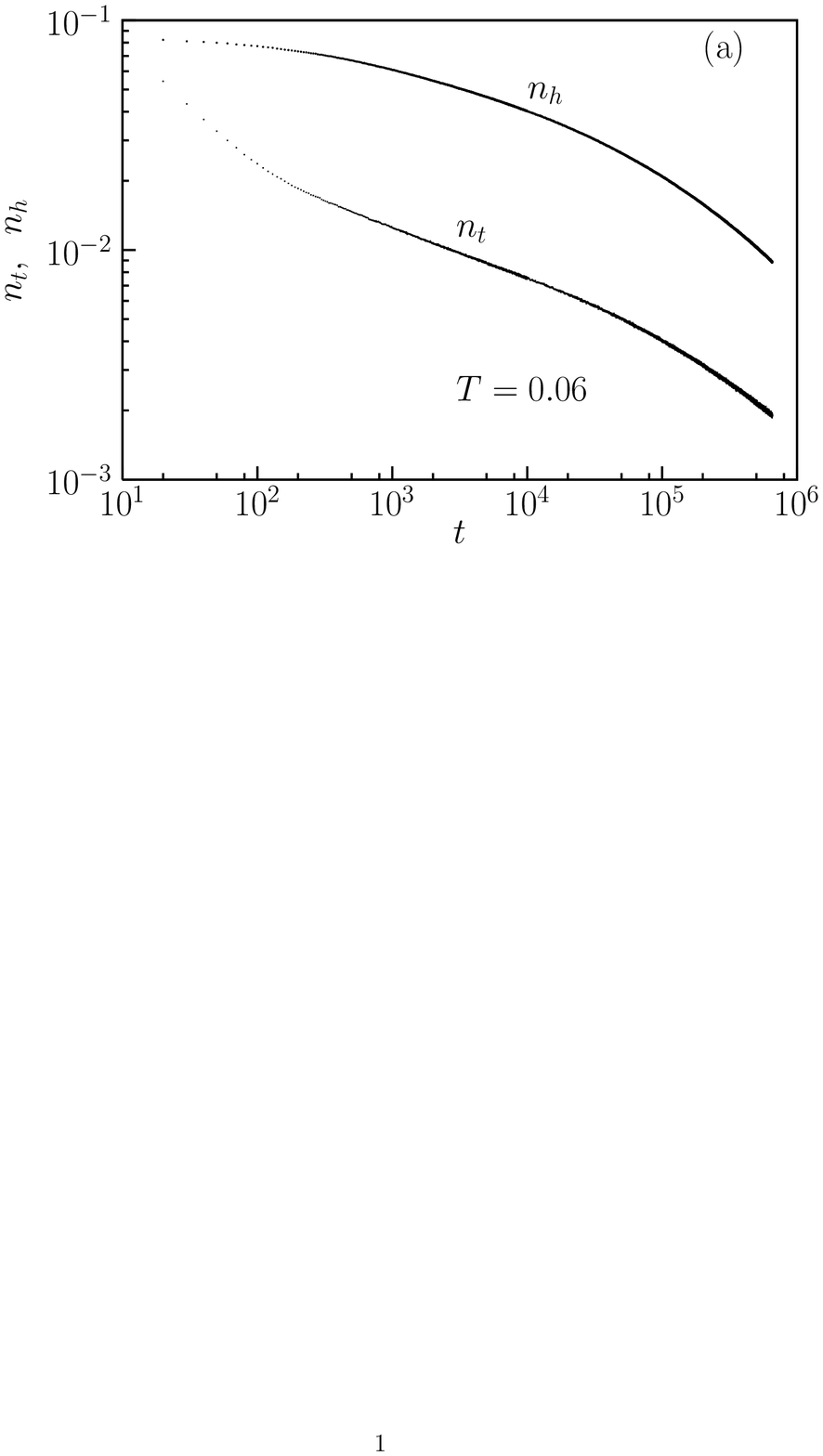}
\vspace*{0.1cm}
\includegraphics[width=8cm]{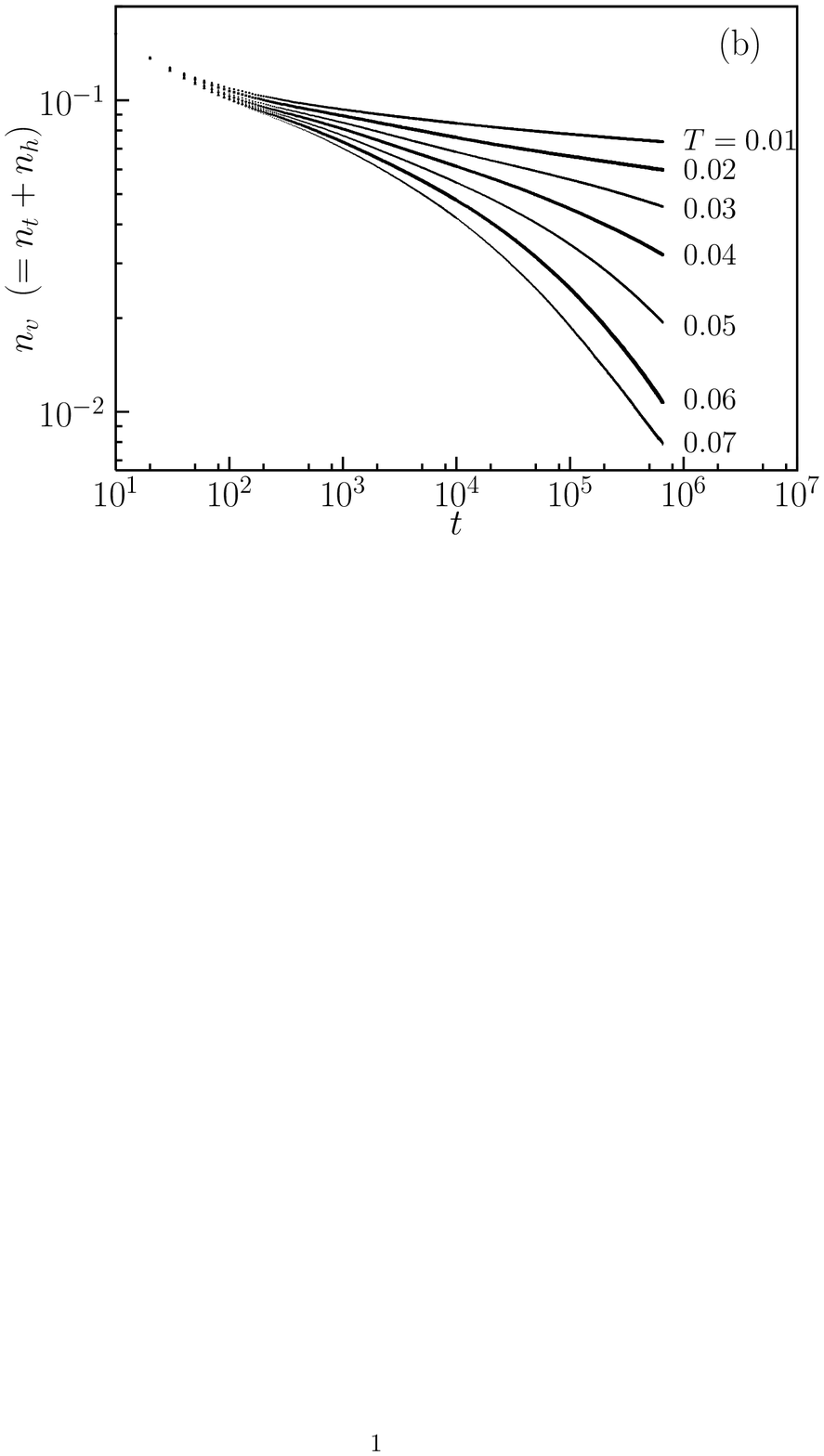}
\vspace*{0.1cm}
\includegraphics[width=8cm]{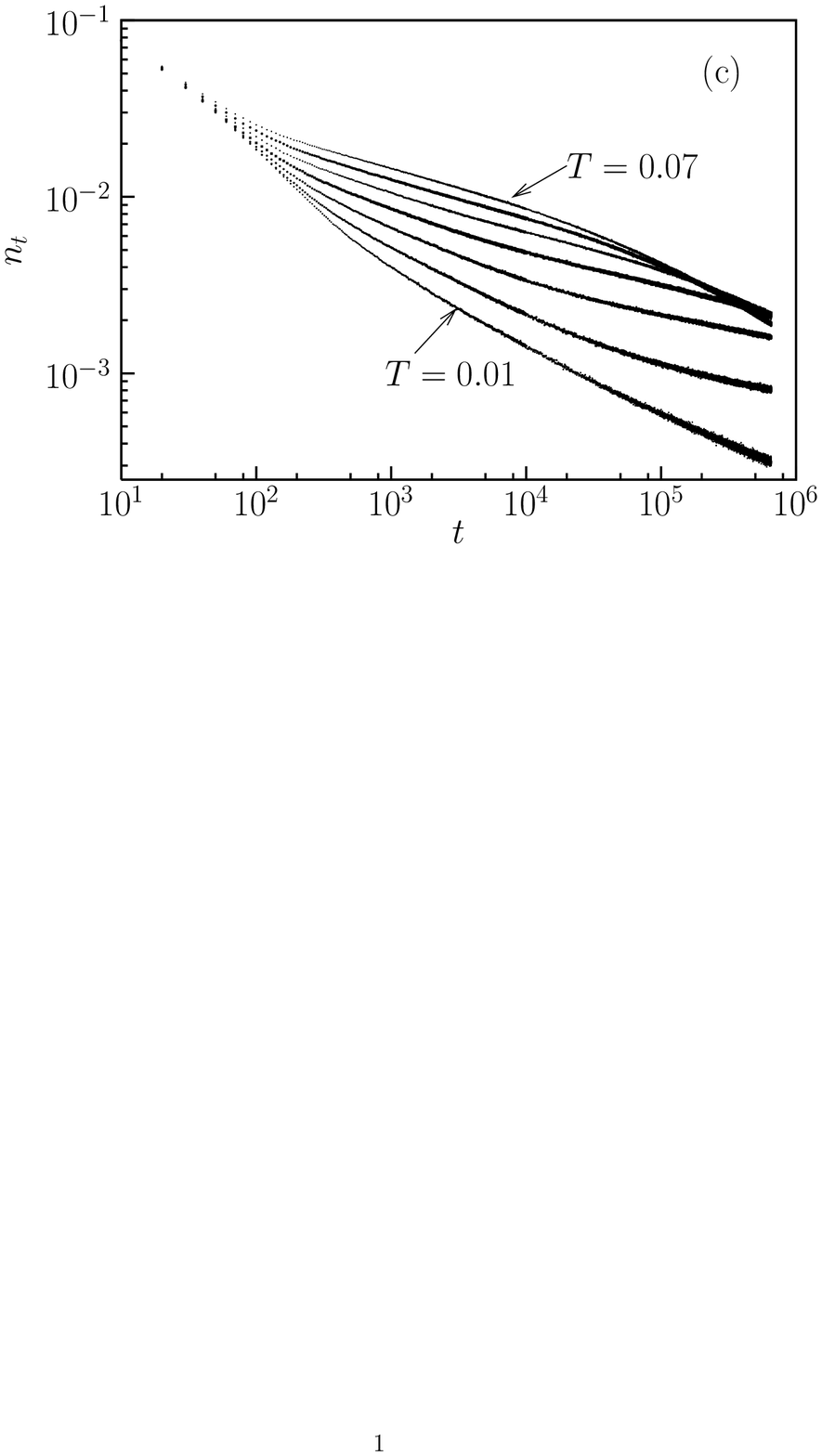}
\vspace*{0.1cm}
\includegraphics[width=8cm]{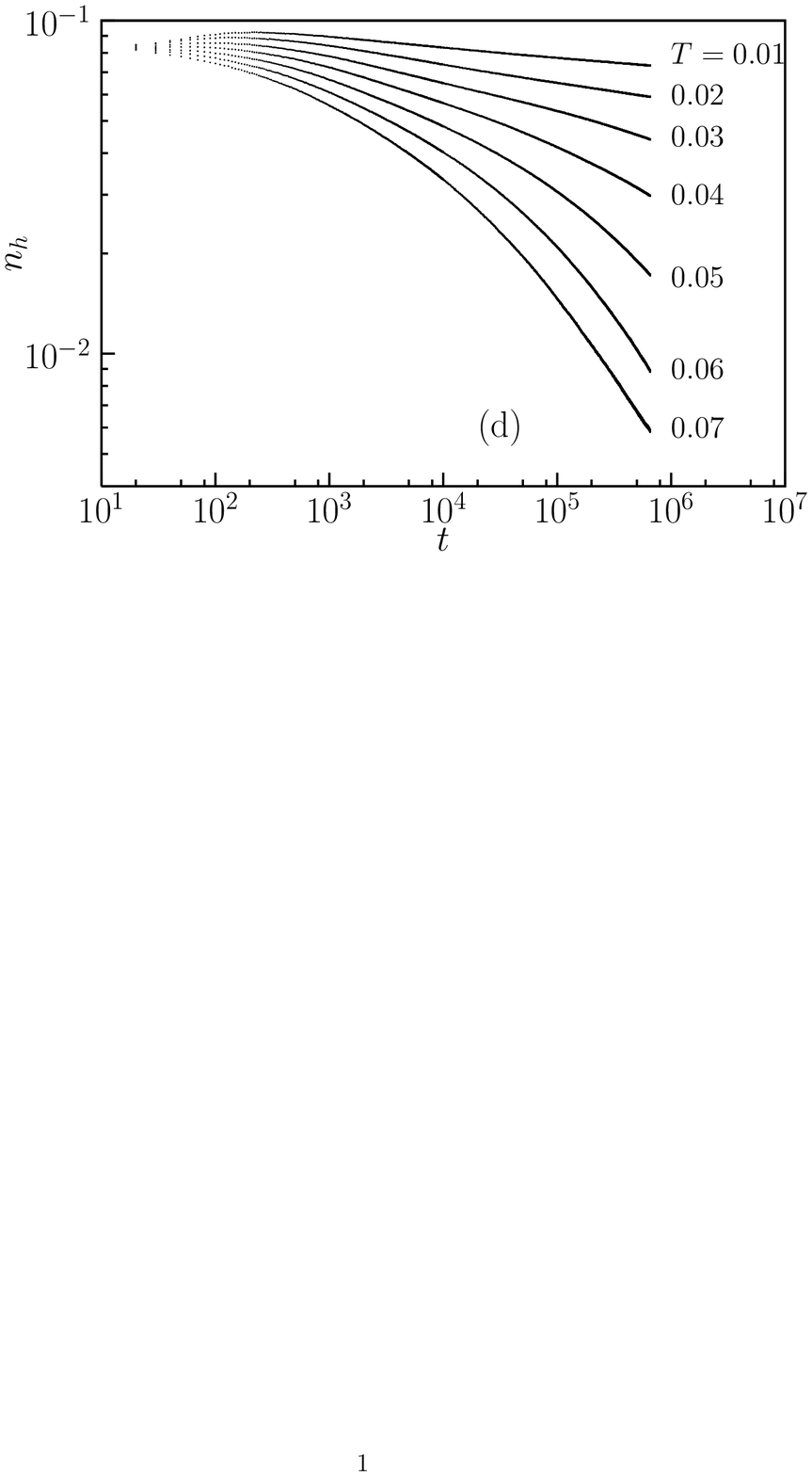}

\caption{(a) $3\theta$ fractional vortex (defect) density vs. time $t$ on hexagonal and 
     triangular plaquettes respectively for $T=0.06$, (b) the total vortex number density (both on 
     hexagonal and triangular plaquettes) vs. time $t$ for $T=0.06$, (c) the time dependence of 
     the density of $3\theta$ vortices on triangular plaquettes at varous temperatures,
     and (d) the time dependence of the density of $3\theta$ vortices on hexagonal plaquettes at 
     varous temperatures. }

\label{defect_number_decay}
\end{figure*}

In order to understand the physical mechanism underlying the breakdown of the critical
dynamic scaling as well as the peculiar time-dependence of the length scale, we 
investigated the time dependence of the total number of the relevant topological defects,
i.e., the fractional $3\theta$-vortices residing on the triangular plaquettes as well as
on the hexagonal plaquettes of the kagome lattice. The results are shown in 
Figs.~\ref{defect_number_decay}(a)-(d) where we can see that, for a given temperature, 
the triangular vortices are decaying much faster than the hexagonal vortices. 
We also find that in general (for both types of the vortices) the decay of the 
total vortex numbers do not exhibit a simple power law behavior valid for the whole 
time range. 
One prominent feature in terms of the temperature dependence of the decay of 
the fractional vortex density is that, for the vortices 
on the triangular plaquettes, the decay rate increases as the temperature is lowered,
while on the other hand, those vortices on the hexagonal plaquettes exhibit opposite
dependence on the temperature with the decay rate decreasing sharply as the temperature
is lowered.  

This can be interpreted as implying that there is almost no (free) energy barrier 
for the motion and decay of the triangular vortices but that some finite barrier exists 
for the hexagonal vortices. We also find an interesting feature of the hexagonal vortices 
(at lower temperatures) where, at initial stage, they increase a little and then start 
to decay. This can probably be understood as the influence of the decay of the neighboring 
triangular vortices with their excess energy turned over to neighboring hexagonal 
vortices, thus generating hexagonal vortices. 
   
Therefore, we can conclude that, for wide range of temperatures, the relaxation of 
the fractional vortices residing on the hexagonal plaquettes determines the coarsening
process such as the correlation length scale corresponding to the quasi-ordered domains.         
In addition, the vortex relaxations exhibit a considerable deviation from power law 
behavior with slowly increasing (in absolute magnitude) local logarithmic slope in 
the late time regime. Especially in the case of the hexagonal vortices, the decay of 
the vortex number density could be fit to a form with 
\begin{eqnarray} \label{eq:vortex_logp}
N_h & = & N_0 \exp [ - b ( \ln (t))^\alpha ] \\
   &  =  & N_0 \exp \left [ - \left (\frac{\ln (t)} {\ln(t_0 )} \right )^\alpha \right ].
\end{eqnarray}
with $b \equiv \ln(t_0 )^{-\alpha}$.
Here, we found that typically $\alpha$ takes values  around $3.3\sim 4.0$ 
(figure not shown here\cite{swpark_prep}).
We found that it could also be fitted to an stretched exponential  form as
\begin{equation} \label{eq:vortex_stexp}
N_h (t) \sim N_0 \exp[-(t/t_0 )^{\alpha}],
\end{equation}
with $\alpha \simeq 0.226 \pm 0.03$.

It appears that the breaking of the critical dynamic scaling is related
to the anomalous relaxation of the number of fractional vortices on the triangular and 
the hexagonal plaquettes. In order to check this possibility, we analyzed the 
behavior of the length scales determined by the vortex distributions as follows.
We define the length scale $\xi_t \equiv  1/\sqrt{n_t}$ which corresponds to the 
average separation between the vortices on the trianular plaquettes.
Similarly, we can define the length scale $\xi_h \equiv  1/\sqrt{n_h}$ corresponding 
to the average separation between vortices on the hexagonal plaquettes.  
We also define the length scales $\xi_v \equiv 1/\sqrt{n_v} = 1/\sqrt{n_h + n_t}$
which represents the length scale corresponding to the total number density of vortices. 
      
\begin{figure*}[htb]
\centering
\includegraphics[width=8cm]{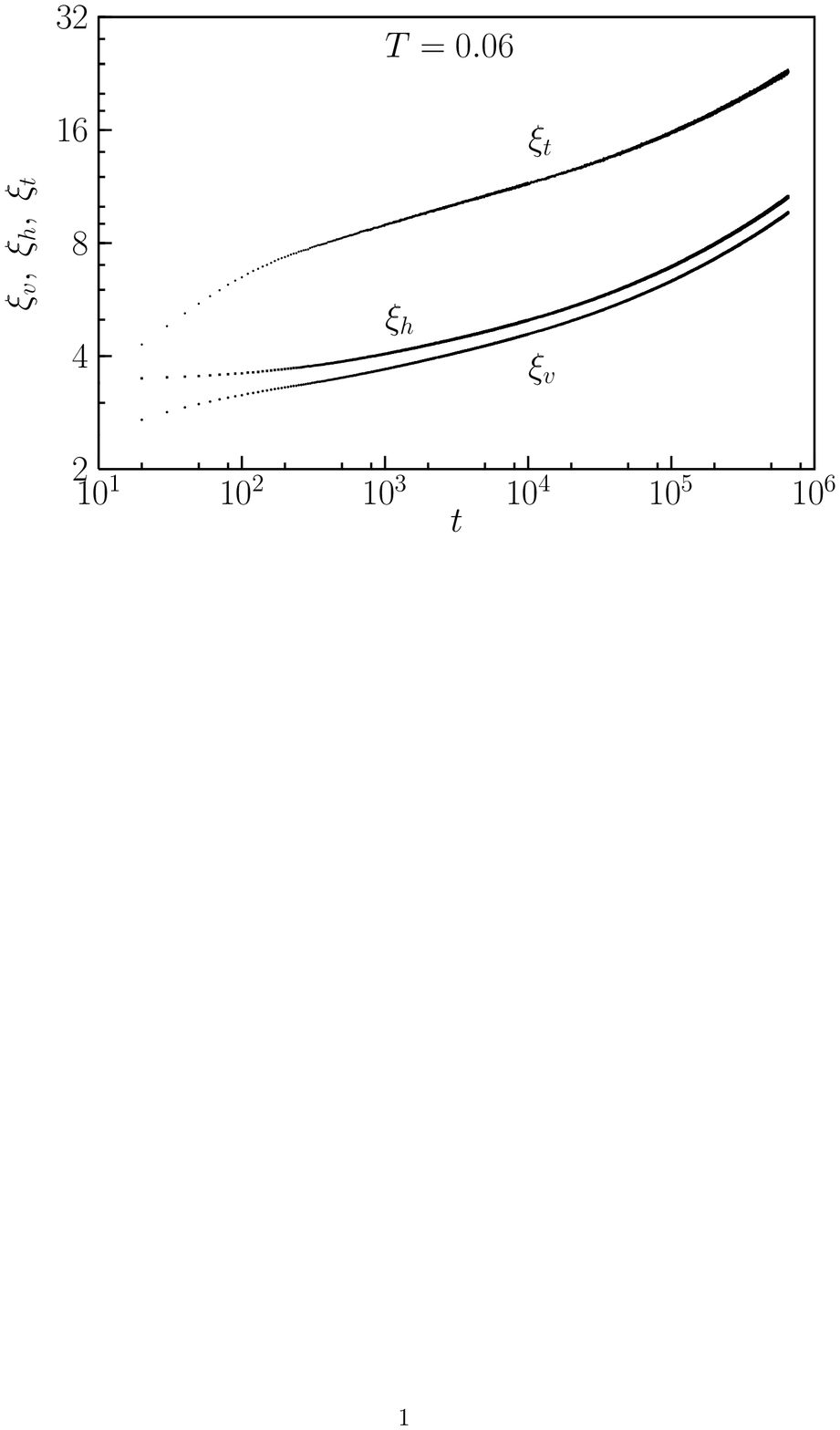}

\caption{The length scales $\xi_h$, $\xi_t$ and $\xi_v$ derived from the vortex densities 
     versus time $t$ for hexagonal, triangular plaquettes and the sum of these, respectively 
     at $T=0.06$. }

\label{defect_length}
\end{figure*}

\begin{figure*}[htb]
\centering
\includegraphics[width=8cm]{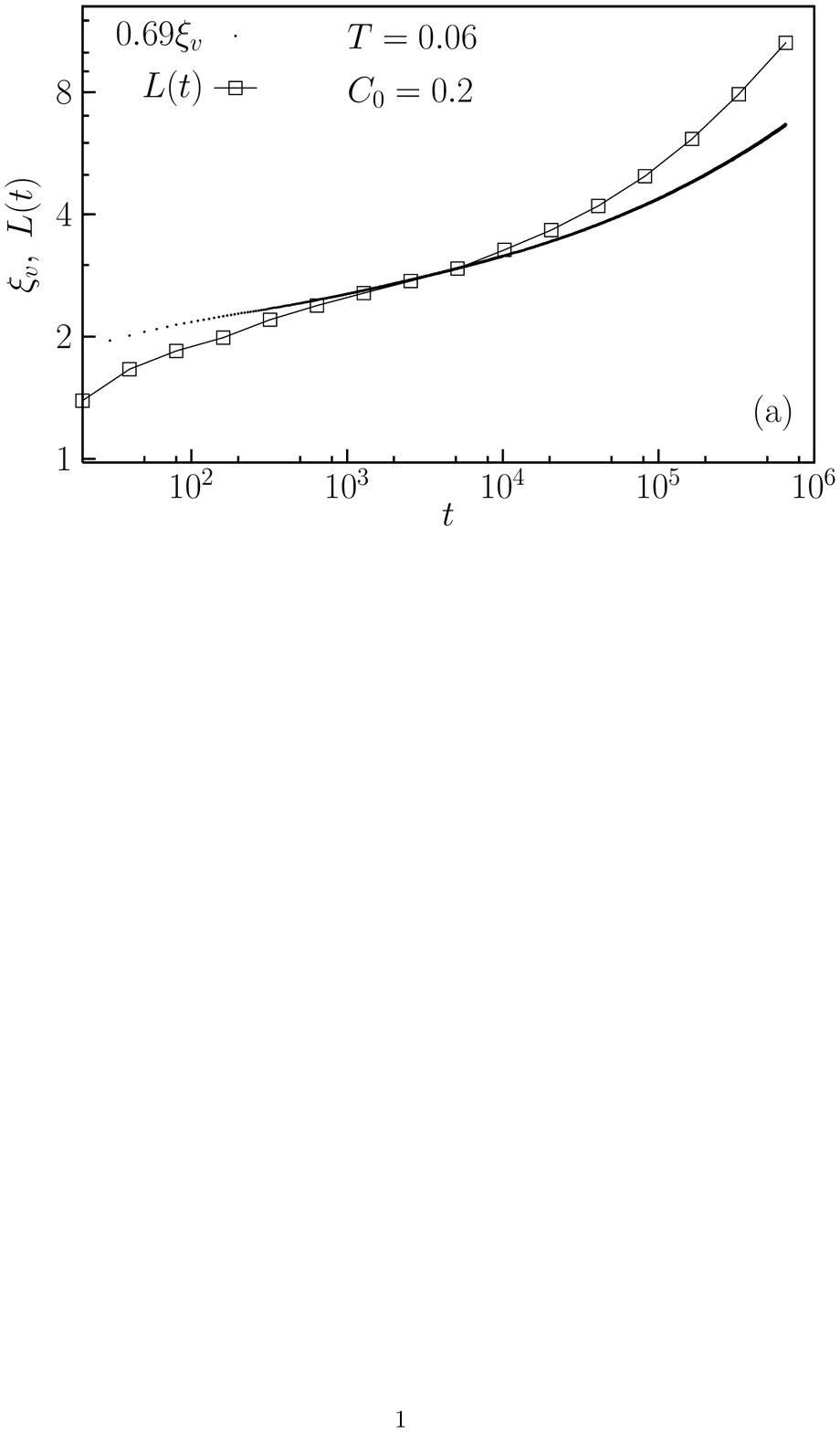}
\vspace*{0.1cm}
\includegraphics[width=8cm]{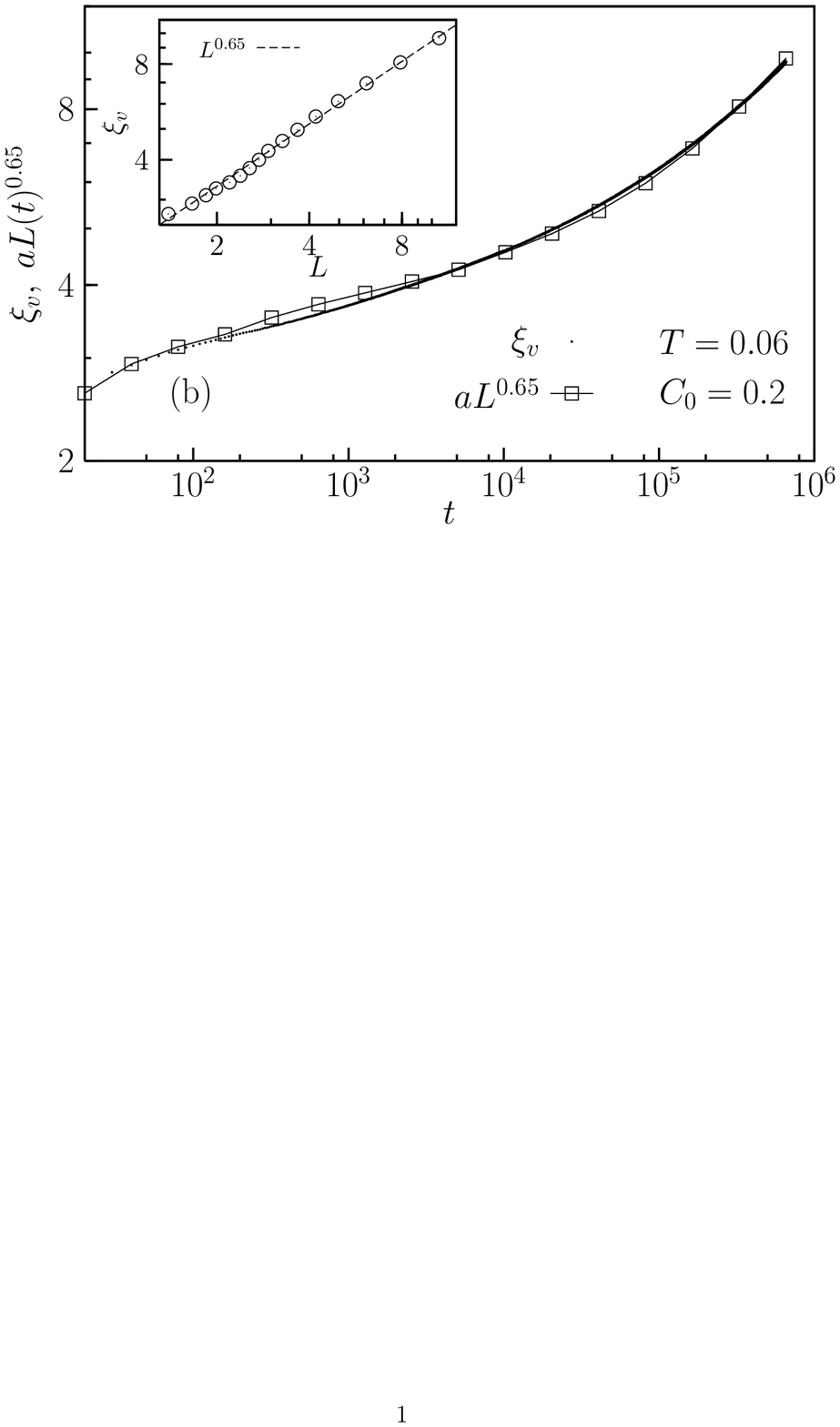}
\vspace*{0.1cm}
\includegraphics[width=8cm]{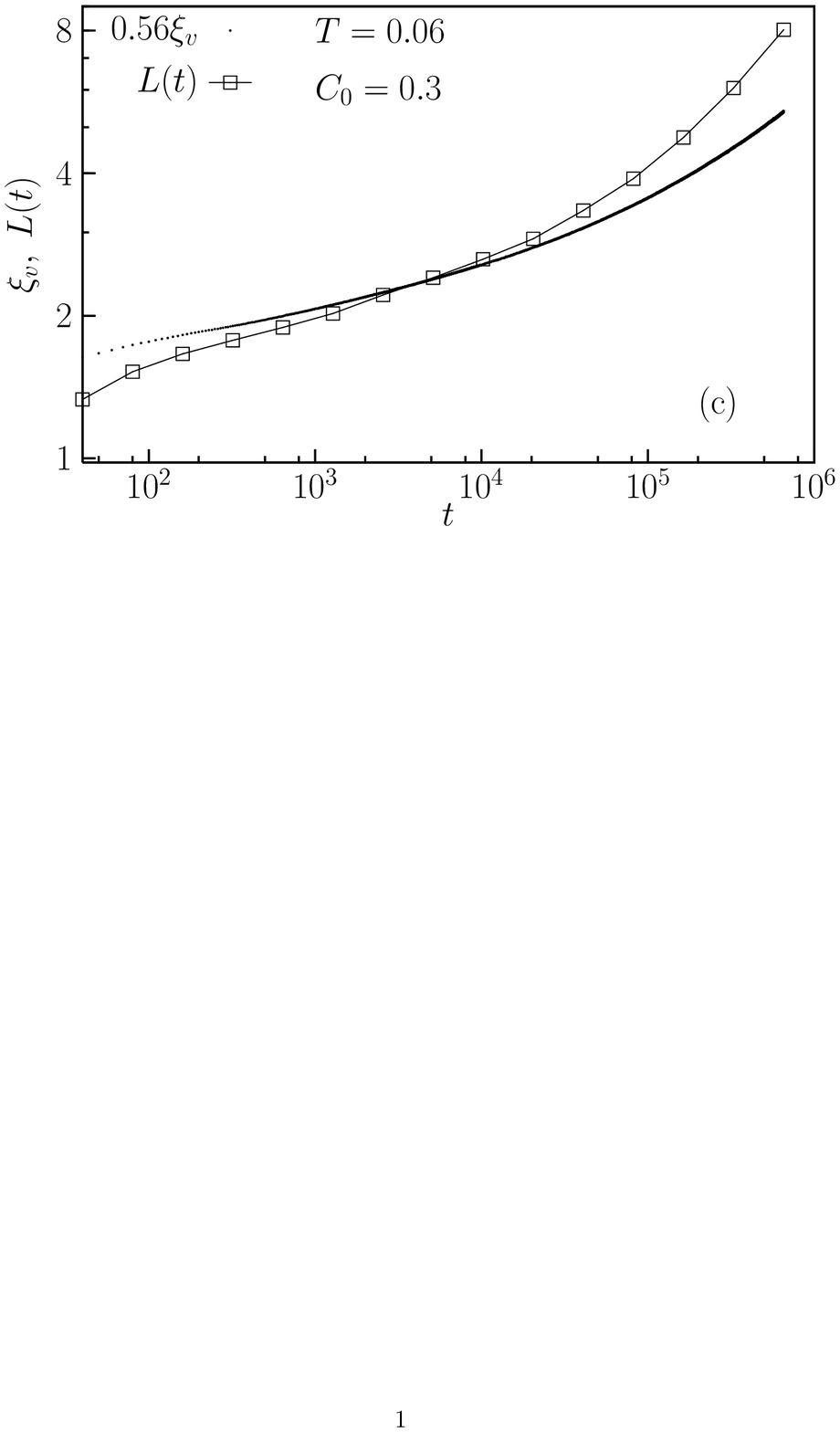}
\vspace*{0.1cm}
\includegraphics[width=8cm]{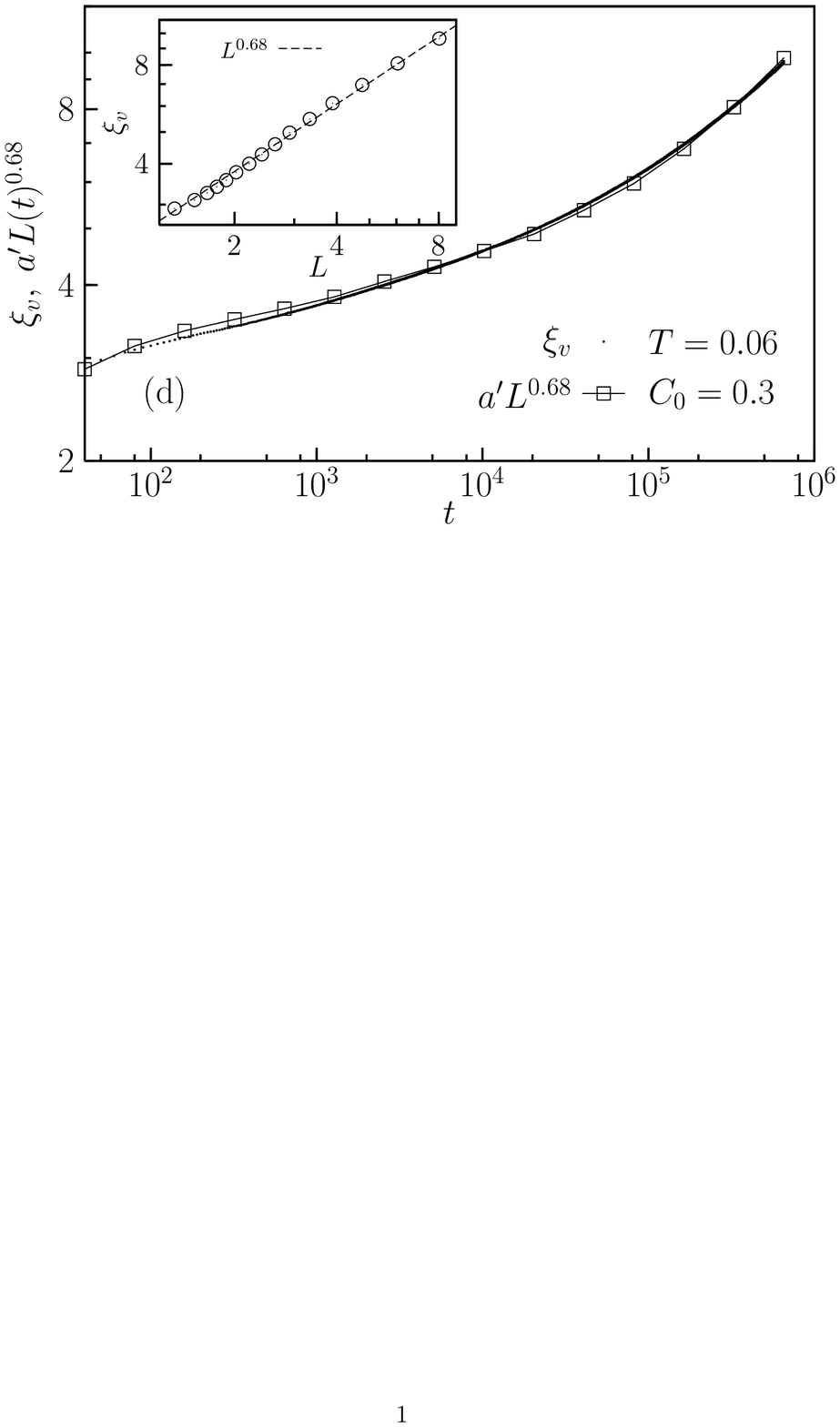}

\caption{ (a) The length scale $\xi_v$ derived from the total vortex density 
     is compared with the domain length scale $L(t)$ obtained from the equal-time
     spatial correlations with $C_0 =0.2$ at $T=0.06$, where one can see that 
     the two length scales are not proportional to each other,  
    (b) a comparison of $L(t)^{0.65}$ (multiplied by a constant factor) 
      with $\xi_v$ which shows an approximate collapse (also see the inset for
      direct comparison between $\xi$ vs. $L$),
    (c) same as (a) with $C_0 = 0.3$, (d) same as (b) with $C_0 = 0.3$ where
     a comparison of $L(t)^{0.68}$ (multiplied by a constant factor) 
      with $\xi_v$ is shown with reasonable collapse (also see the inset).  
    (Note that in (a) and (c) $\xi_v$ is multiplied by a constant for convenience of 
    comparison.) }

\label{growth_vs_defect_length}
\end{figure*}

Figure~\ref{defect_length} shows the three length scales at $T=0.06$ which clearly shows 
the faster growth of the length scale for the triangular vortices compared with 
that for the hexagonal vortices. In Fig.~\ref{growth_vs_defect_length}(a)-(d) comparison 
is made between the length scales $L(t)$ derived from the spatial correlation functions
and the length scales $\xi_v$ derived from the total vortex densities. We see that
(for both cases of $C_0 =0.2$ and $C_0 = 0.3$) there exists no simple proportional 
relationship between the two length scales. 
Rather, the length scale $L(t)$ from the spatial 
correlation is seen to grow faster than the vortex lenth scale $\xi_v$. 
This probably implies that the vortices are not distributed evenly (statistically speaking)
in the system but that the vortices are somehow distributed in a non-random manner 
such that some degree of clustering occurs. Interesting result 
is that the two length scales satisfy some nontrivial relationships such that 
$\xi_v \sim L(t)^{\lambda}$ with $\lambda \simeq 0.65$ for the case of
 $C_0 =0.2$ (Fig.~\ref{growth_vs_defect_length}(b))
and $\lambda \simeq 0.68$ for the case of $C_0 =0.3$ (Fig.~\ref{growth_vs_defect_length}(d)). 
This means that the vortex configuration of the system do not exhibit simple 
self-similarity at different time instants. Rather the vortices probably tend to cluster
more unevenly as time passes by, leading to the correlation length scale $L(t)$ growing
faster than the length scale derived from the vortex density.   

\begin{figure*}[htb]
\centering
\includegraphics[width=8cm]{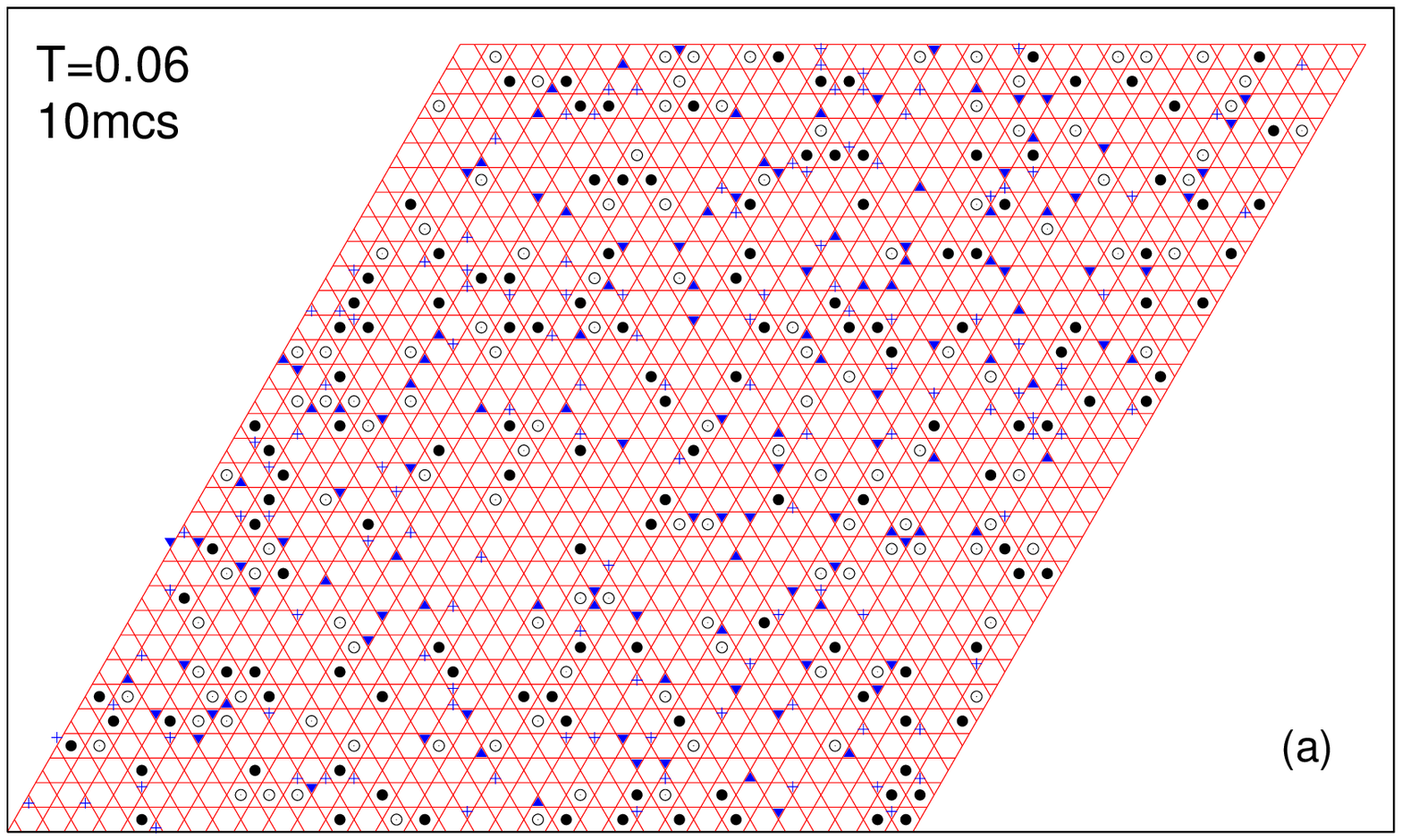}
\vspace*{0.1cm}
\includegraphics[width=8cm]{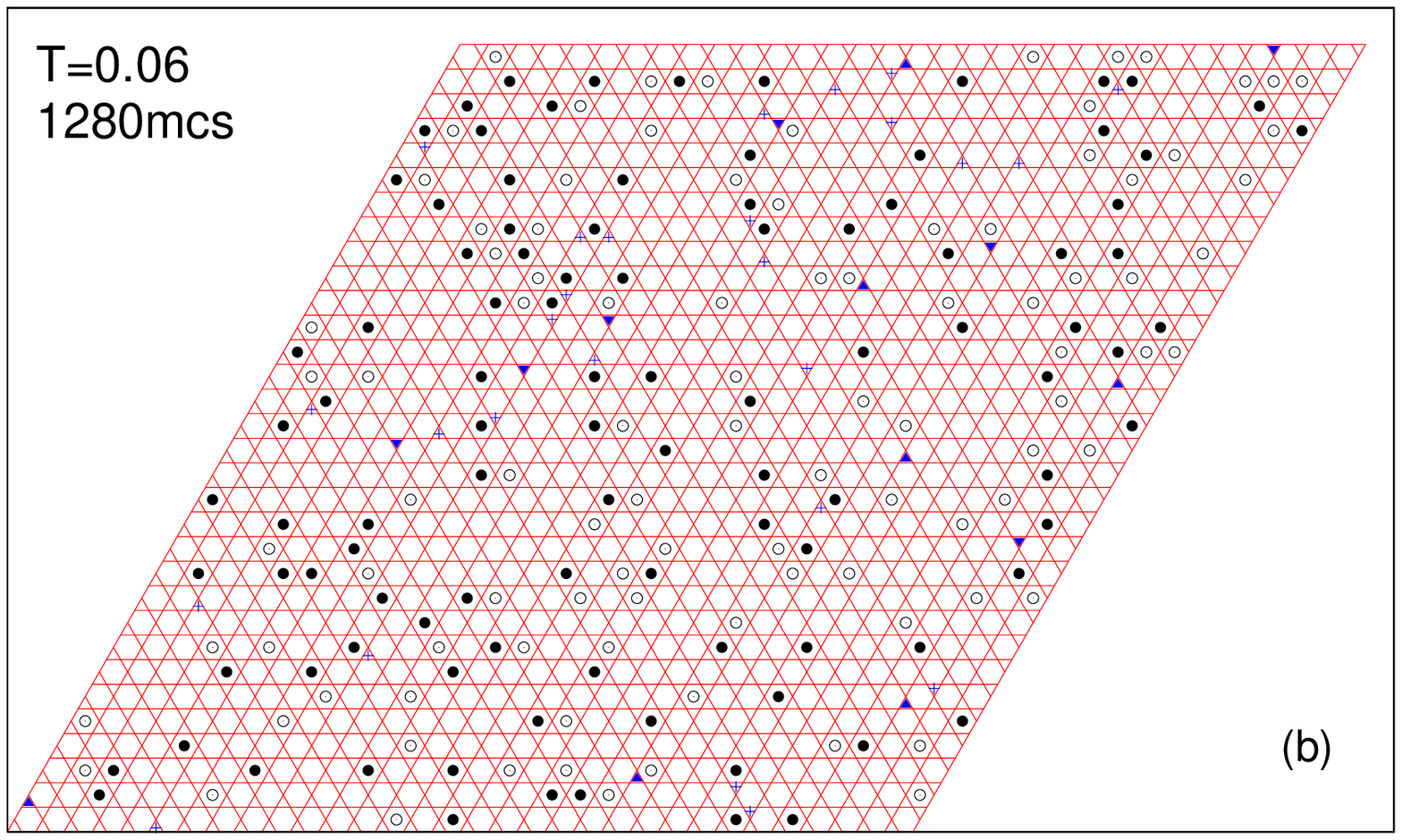}
\vspace*{0.1cm}
\includegraphics[width=8cm]{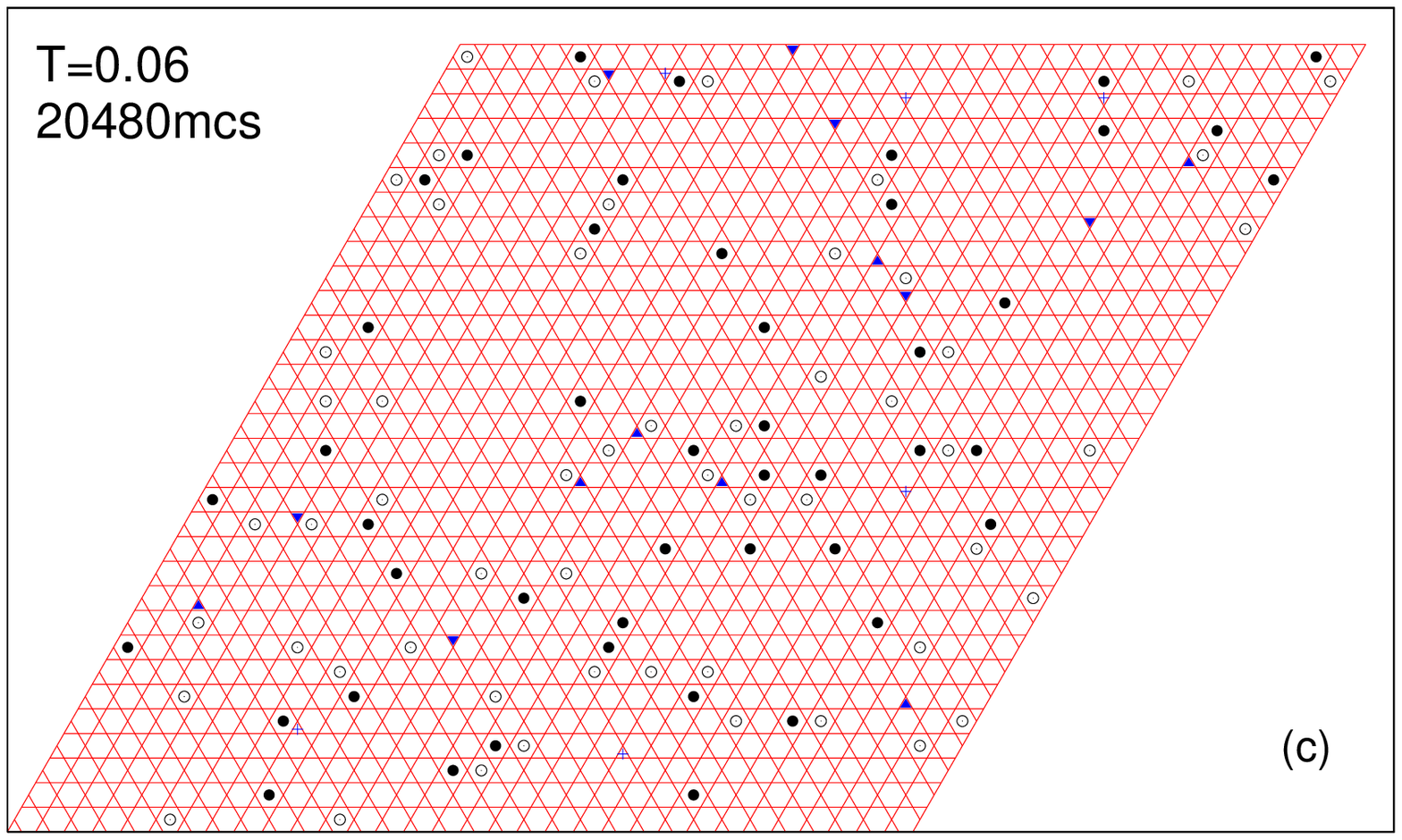}
\vspace*{0.1cm}
\includegraphics[width=8cm]{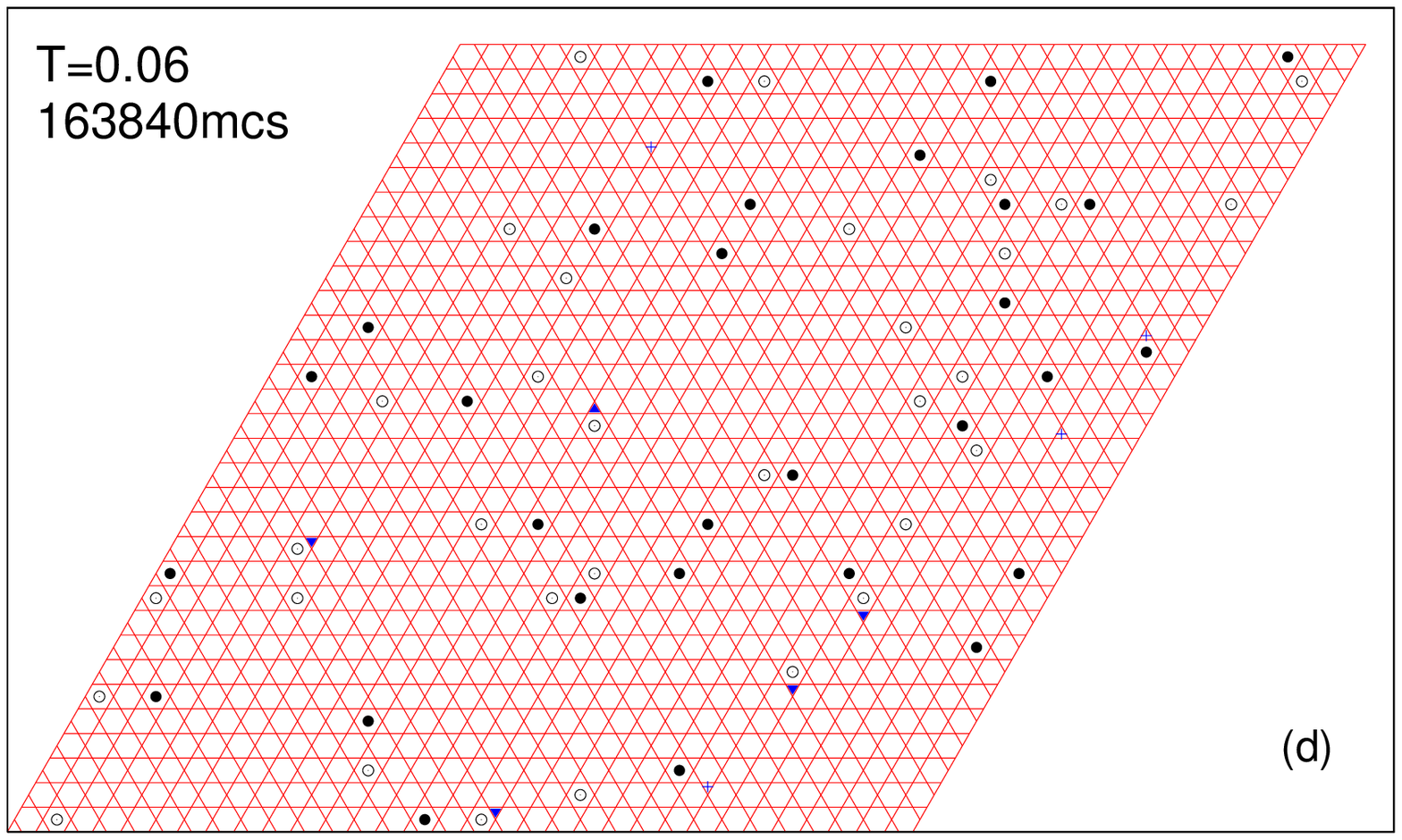}

\caption{(Color online) Configurations of $3\theta$-vortices at different time steps, for 
  (a) $t=10$ MCS (b) $t=1280$ MCS, (c) $t =20480$ MCS, and $t=163840$ MCS. Filled (empty) 
   circles represent postive (negative) vortices on hexagonal plaquettes and 
   pluses represent positive vortices on the triangular plaquettes while filled triangles 
   negative vortices (on the triangular plaquettes).}   
\label{config}
\end{figure*}

The snapshot of defect configurations  for the case of $T=0.06$ is shown in 
Fig.~\ref{config}(a)-(d). We can easily recognize faster decay of those vortices 
on the triangular plaquettes.
From the snapshots alone, however, it is not easy to detect the tendency of relative clustering 
of the vortices in the late time regime.

Now we turn to the autocorrelation functions of the variables $\exp(i\theta)$ and 
$\exp(3i\theta)$. For these two variables, we find that the simulation results on 
the autocorrelations are very different. For the case of $\exp(i3\theta)$, we 
obtain a approximately power-law decay behavior $A_{3}(t) \sim t^{-\lambda}$ in the early
time regime up to around $t \simeq 10^3$.
The value of the exponent $\lambda$ ranges from $\lambda \simeq 0.56$ (for $T=0.01$) 
to $\lambda \simeq 0.90$ (for $T=0.07$). 
The exponent tends to increase slightly at higher temperatures. At later time, 
however, the autocorrelations decay faster than power-law relaxation of the early
time stage.  Due to the larger statistical fluctuations in the relaxation of
the autocorrelation $A_{3}(t)$, we could not attempt a comparison of $A_{3}(t)$ with
the growing length scale $L(t)$.    

\begin{figure*}[htb]
\centering
\includegraphics[width=8cm]{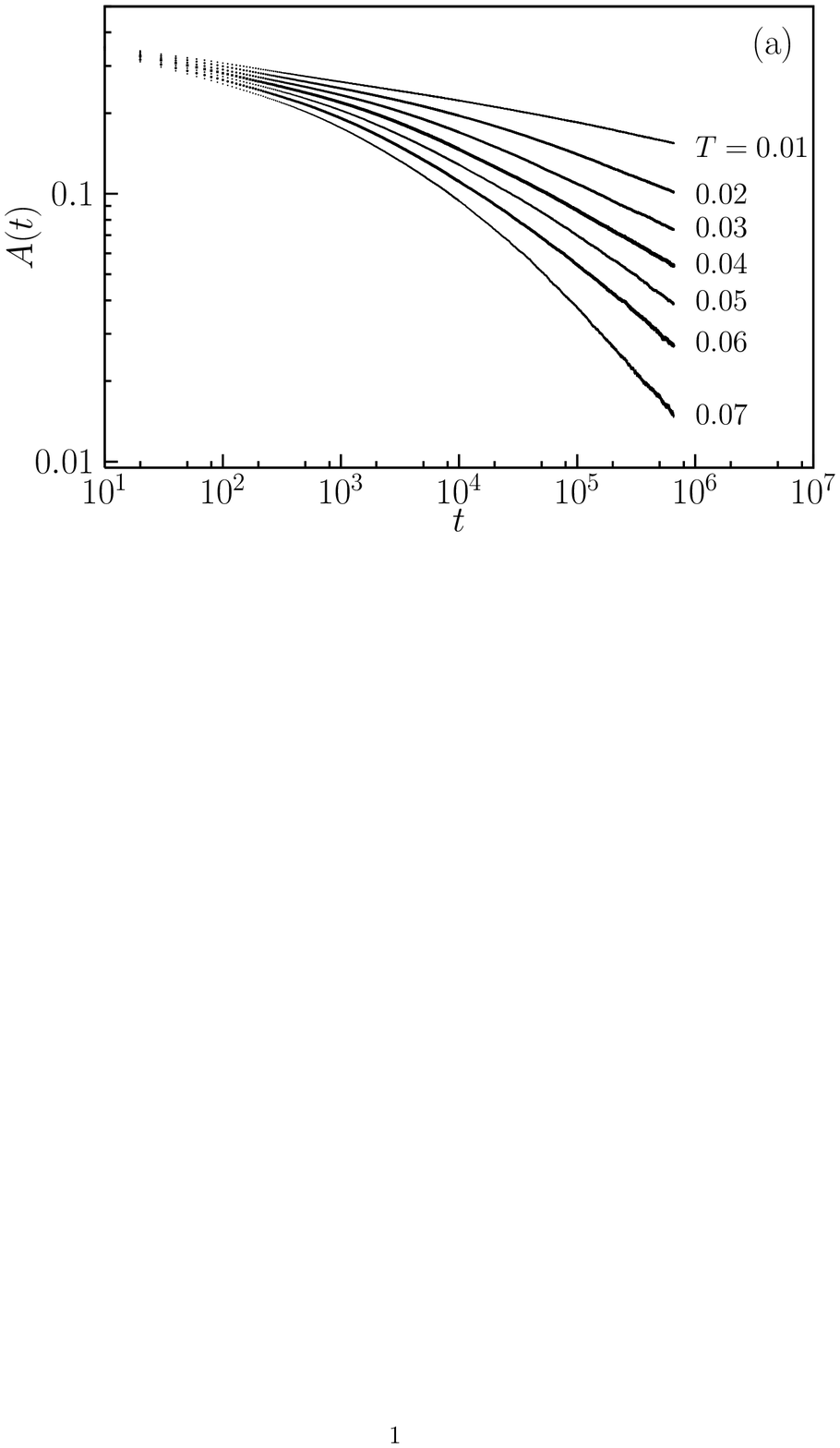}
\includegraphics[width=8cm]{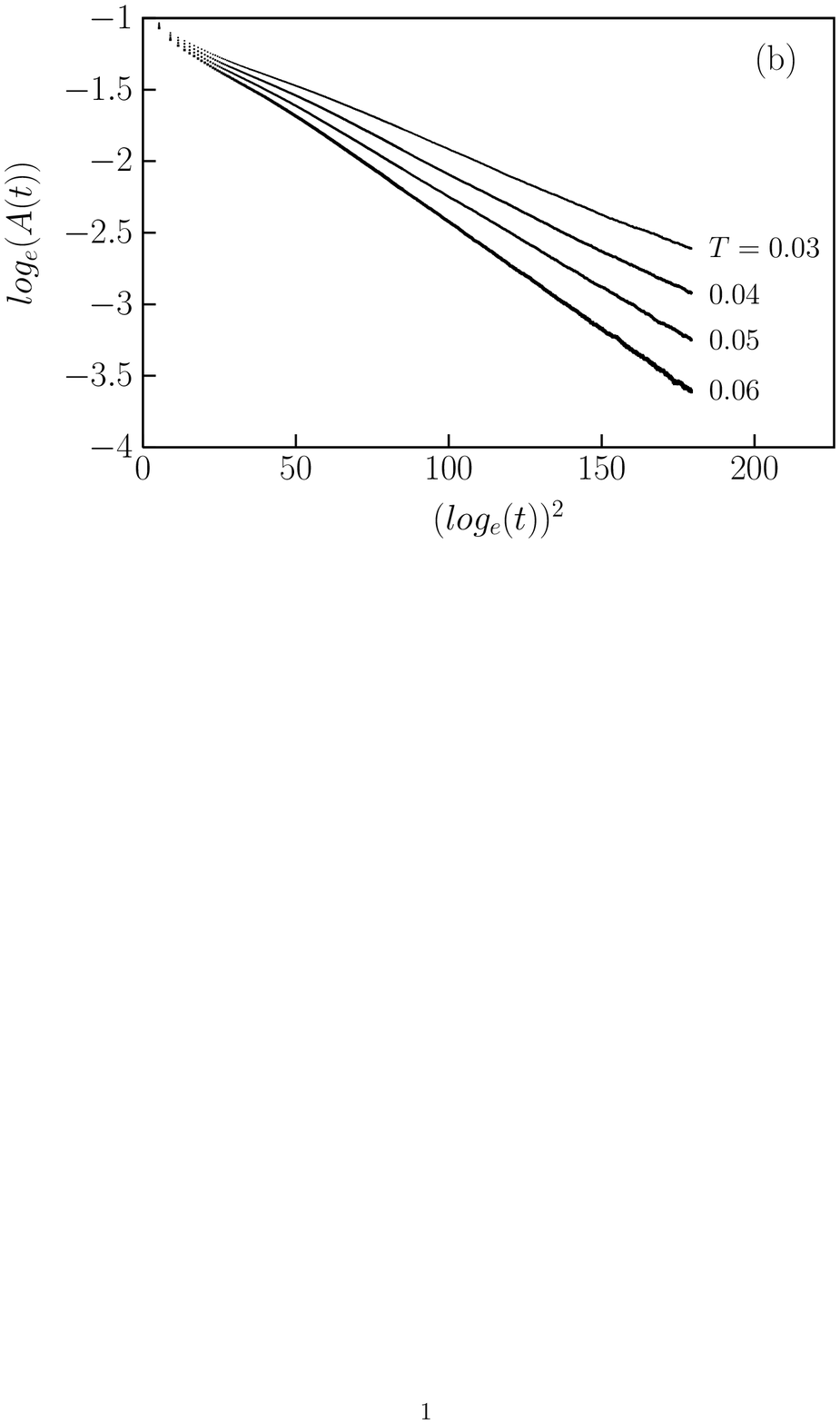}

\caption{(a) Autocorrelation functions for $\exp(i\theta)$ for different
temperatures, where we can discern clear signs of faster-than-power-law decay,
(b) plots for the logarithm of the autocorrelation functions for $\exp(i\theta)$ vs. 
 $(\log(t))^{2}$ which show approximate linear relationships.}
\label{autocorr}
\end{figure*}

In contrast, in case of the phase $\exp(i\theta)$, the spin autocorrelaton 
exhibits less statistical fluctuations with a non-power law behavior\cite{cherepa94} 
that can be reasonably fitted by
\begin{eqnarray} \label{eq:autoco_th}
A(t) & = & A_0 \exp[-b(\ln(t))^{\gamma}] \\
    &  = & A_0 \exp \left [- \left ( \frac{\ln(t)}{\ln(t_0 )}\right )^{\gamma} \right ]
\end{eqnarray}
with the exponent $\gamma = 2.0 \pm 0.3$ and $b \equiv \ln(t_0 )^{-\gamma}$.
One of the simulation results is shown in Fig.~\ref{autocorr}(a-b) for the case 
of $T=0.06$ with $\widetilde N$, which shows a suitable fit to the above functional 
form with $b \simeq 0.15$ and $\gamma \simeq 2.0$. Note that, in the limit of 
$\gamma =1$, $A(t)$ reduces to a power law of $A(t) \sim A_0{t}^{-b}$. 
The fitted values of $\gamma$ implies that $A(t)$ exhibits a considerable 
deviation from a power-law behavior. 
We also attempted to fit the simulation results to stretched exponential form.
This resulted in rather small values for the stretching exponent of 
$\alpha \simeq 0.14$. 
It is interesting to note that similar behavior of relaxation in the autocorrelation 
function of the order parameter was reported in the coarsening dynamics of the 
so-called Hamiltonian {\em XY} model on a square lattice.\cite{kjkoo2006} 

In summary, we investigated the coarsening dynamics of the antiferromagnetic {\em XY} 
model on a kagome lattice. We found that the critical dynamic scaling is violated.
Novel and unusual coarsening dynamics may be attributable to the existence of
infinite ground state degeneracies leading to nontrivial decay behavior of 
the density of defects residing on the triangular and hexagonal plaquettes
with nontrivial growth of multiple length scales. The competition between the critical 
thermal fluctuations of the equilibrium and the infinite ground state degeneracy
probably gives rise to the nontrivial features of the relaxation dynamics.
It may be possible to collapse the equal-time spatial correlation functions 
through a multiscaling scheme which we haven't tried yet.   
It would be also interesting to investigate the aging dynamics of this model
system.\cite{wills2000}

\newpage

\newpage

\newpage

\begin{figure*}[htb]
\centering
\includegraphics[width=8cm]{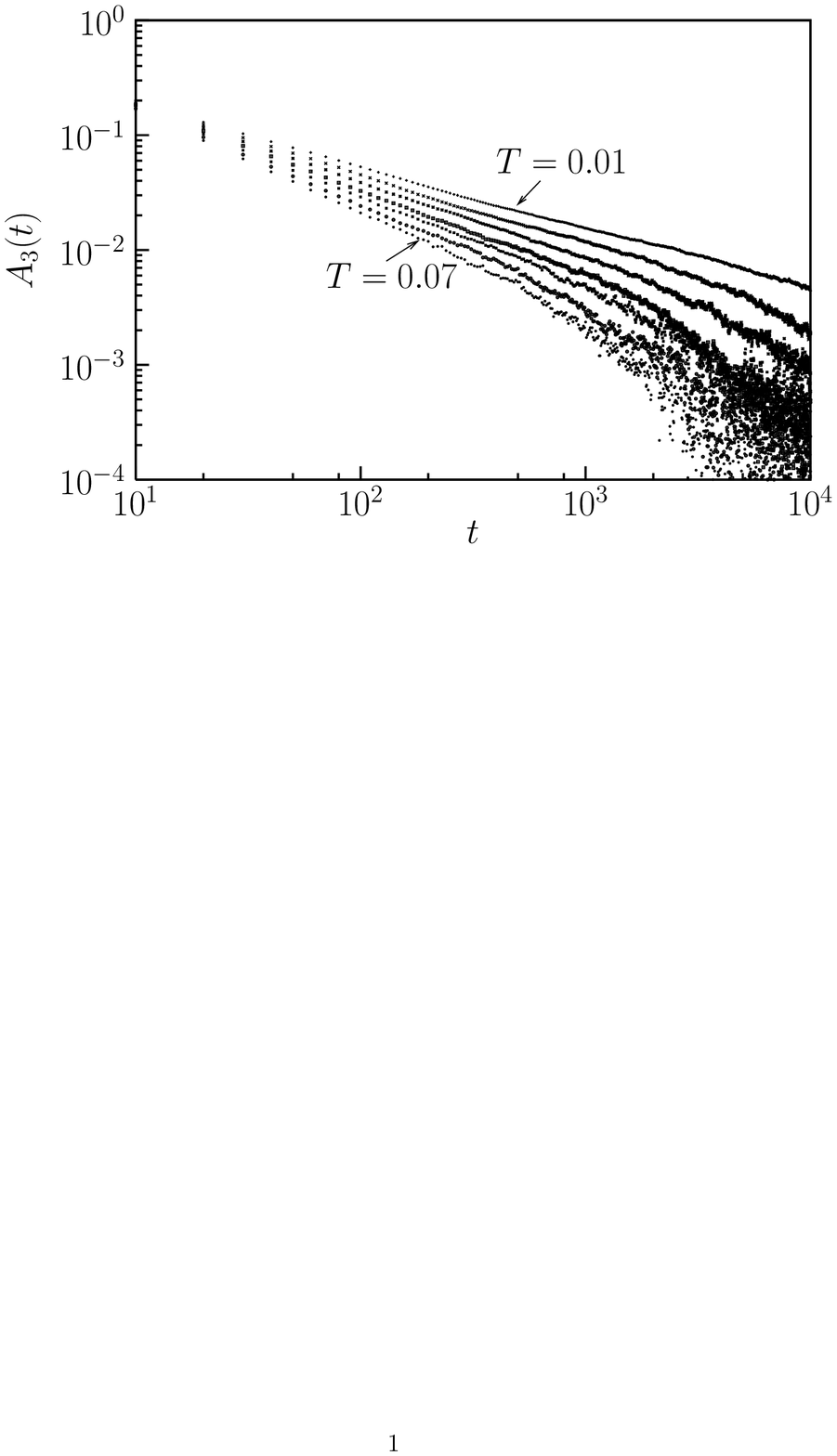}

\caption{Autocorrelation functions for $\exp(3i\theta)$ for different
temperatures. We can see that the autocorrelations exhibit approximate power law 
 decay up to intermediate time stages and then faster decay in the late time stage. }
\label{autocorr3}
\end{figure*}

\end{document}